\documentclass[useAMS,usenatbib]{mn2e}
\usepackage[pdftex]{graphicx}
\newcommand\msun{\rm{M_{\odot}}}

\def\stacksymbols #1#2#3#4{\def\theguybelow{#2}
        \def\verticalposition{\lower#3pt}
        \def\spacingwithinsymbol{\baselineskip0pt\lineskip#4pt}
        \mathrel{\mathpalette\intermediary#1}}
\def\intermediary #1#2{\verticalposition\vbox{\spacingwithinsymbol
        \everycr={}\tabskip0pt
        \halign{$\mathsurround0pt#1\hfil##\hfil$\crcr#2\crcr
                \theguybelow\crcr}}}

\def\lta{\stacksymbols{<}{\sim}{2.5}{.2}}

\title[3D simulations: gently heating a cool group]
{AGN Feedback in Galaxy Groups: \\ 
the Delicate Touch of Self-Regulated Outflows}

\author[M. Gaspari et al.]{M. Gaspari$^{1}$\thanks{E-mail:
massimo.gaspari4@unibo.it},
F. Brighenti$^1$,
A. D'Ercole$^2$, C. Melioli$^1$ \\
$^{1}$Astronomy Department, University of Bologna, Via Ranzani 1, 
40127 Bologna, Italy\\
$^{2}$INAF-OABO, Via Ranzani 1, 40127 Bologna, Italy
}

\begin{document}

\date{Accepted 2011 March 25.  Received 2011 March 25; in original form 2011 February 8}

\pagerange{\pageref{firstpage}--\pageref{lastpage}} \pubyear{2011}

\maketitle

\label{firstpage}


\begin{abstract}
AGN heating, through massive subrelativistic outflows, might be
the key to solve the long-lasting `cooling flow problem' in
cosmological systems. In a previous paper, we showed that cold accretion
feedback and, to a lesser degree, Bondi self-regulated models are in fact able 
to quench cooling rates for several Gyr, at the same time preserving the main
cool core features, like observed density and temperature profiles. 
Is it true also for lighter systems, such as galaxy groups?
The answer is globally yes, although with
remarkable differences. Adopting a modified version of the AMR code
FLASH 3.2, we found that successful 3D simulations with cold and Bondi 
models are almost
convergent in the galaxy group environment,
with mechanical efficiencies in the range $5\times10^{-4}-10^{-3}$
and $5\times10^{-2}-10^{-1}$, respectively. 
The evolutionary storyline of galaxy groups
is dominated by a quasi-continuous gentle injection with sub-Eddington
outflows (with mechanical power $P\sim10^{44}$ erg s$^{-1}$, 
$v\sim10^{4}$ km s$^{-1}$). The cold
and hybrid accretion models present, in addition, very short quiescence periods, followed by
moderate outbursts (10 times the previous phase), which generate a series
of $10-20$ kpc size cavities with high density contrast, temperatures
similar to the ambient medium and cold rims. After shock heating,
a phase of turbulence promotes gas mixing and diffusion of metals,
which peak along jet-axis (up to 40 kpc) during active phases. At this stage the tunnel,
produced by the enduring outflow
(hard to detect in the mock X-ray surface brightness maps),
is easily fragmented, producing tiny buoyant bubbles, typically a few kpc in size.
In contrast to galaxy clusters, the AGN self-regulated feedback has to be
persistent, with a `delicate touch', rather than rare and explosive strokes.
This evolutionary difference dictates 
that galaxy groups are not scaled-down versions
of clusters: AGN heating might operate in different regimes,
contributing to the self-similarity breaking observed. 
\end{abstract}


\section[]{Introduction}

Remarkably, most of the galaxies in the universe reside in small groups
(up to 70\% in the nearby zone; \citealt{geh83,mul00}). 
Nevertheless, relatively little attention has been
devoted to groups with respect to
their `big brothers', galaxy clusters.
Groups, with typical masses 
around $10^{13}\msun$, are thus a key class of objects,
containing a significant fraction of the overall universal baryon
budget and being a nursery for massive (early-type) galaxies. 

Galaxy groups are
X-ray sources, with diffuse extended (50-500 kpc) emission similar to that of rich clusters,
 but with lower luminosities, ranging from 10$^{41}$ erg s$^{-1}$ 
up to several times 10$^{43}$ erg s$^{-1}$, with temperatures around
$\sim 0.3-2$ keV (\citealt{mul00}).


It is known that clusters and groups do not follow the global
scaling relations predicted by self-similar models  (e.g. 
\citealt{kai86,evr91,osp04,vik06}).
Groups are not perfect scaled down versions of clusters.
If groups are sampled at sufficiently large
radii, some X-ray scaling relations (e.g. luminosity$-$temperature,
mass$-$temperature) seem similar to those
followed by massive clusters (\citealt{osp04,sun09}), with the
scatter increasing for $T\lta 1$ keV.
However, systematic differences between groups and clusters
exist, with the former having clearly lower gas (and baryon) fractions
and flatter entropy profiles (\citealt{fin05,voi05,sun09}). 
Moreover, groups have velocity dispersions of several 
hundreds km s$^{-1}$, comparable to the internal velocity dispersion
of individual galaxies. For this reason, some processes are more
(merging) or less (ram-pressure
stripping, galaxy harassment) relevant than in clusters.


The discrepancies between the observed scaling relations and the
predictions of self-similar models, 
and especially the
systematic differences between groups and clusters quoted above,
can be explained by the presence of non-gravitational heating,
acting during or after the gravitational collapse. In lower mass
systems the same amount of energy per unit mass has a stronger
effect. This is a key point for the mechanical AGN feedback problem investigated
here and in our previous work (\citealt{gas11}, G11 thereafter).


As in clusters and elliptical galaxies, 
many galaxy groups display the typical signatures of cooling flows\footnote{
Interestingly, the analogues of non-cool core clusters on the group scale have
been rarely detected (\citealt{joh11}).}
(a most notable example being NGC 5044). In the center, where
the radiative cooling time is the shortest ($\sim 10^7$ yr), 
the temperature of the intragroup medium (IGM)
drops by nearly 50\%.
Moreover, the X-ray surface brightness is sharply peaked, with the maximum
coincident with a massive elliptical galaxy.

A large sample of X-ray groups is presented in \citet{sun09}.
Gathering 43 objects with {\it Chandra} data, they confirmed 
the existence of a `universal' temperature 
profile (\citealt{dgm02,voi05,lem08}).
The group profiles are slightly steeper than those of clusters for
$r\ge 0.15\;r_{500}$.
In the inner cool core, the scatter is significant: 
the majority presents positive gradients, while few are
approaching a flat temperature profile (a possible signature of a heating cycle). 
Sun et al. produced also the entropy profiles, which retain the thermal history of the
gas.
They again confirmed previous important works 
(\citealt{pon99,pon03,fin02}), suggesting deviations in the entropy 
from the self-similar relation ($K\propto T$),
showing a clear flat core, and an excess 
relative to massive clusters.

One of the most striking riddles of galaxy group astrophysics concerns
the so called `cooling flow problem'. This paper attempts to solve it, finding plausible 
consistent models, as has been done for galaxy clusters in G11.
The classical cooling flow (CF) model (\citealt{fab94} for a review) predicts that, as the gas
radiates, its entropy decreases, and thus gets compressed by the surrounding gas, causing it
to subsonically flow inward. The estimated (isobaric) cooling rates is given by:
\begin{equation}
\dot{M} \simeq \frac{2}{5} \frac{\mu m_{\rm p}}{k_{\rm B} T}L_{\rm X},
\end{equation}
where $L_{\rm X}$ is the X-ray luminosity, $\mu$ the mean molecular weight, $k_{\rm B}$
the Boltzmann constant, $m_{\rm p}$ the proton mass.
For typical luminosities of groups ($\approx 10^{43}$ erg s$^{-1}$), 
the relation implies cooling rates of several tens $\msun$/yr.
Such high predicted $\dot{M}$ have been in contrast for years with the
estimates of
star formation rates and cold gas masses (at least an order of magnitude lower), 
but the definitive observational evidence against the traditional CF
model has been the spectral analysis of XMM-RGS spectra
(e.g. \citealt{pet01,pet03,tam03,pef06}).\\


In the last decade the new generation {\it XMM-Newton} and {\it
Chandra}  telescopes have
radically changed the panorama of the evolution of hot gas in groups
and clusters.
{\it ROSAT-HRI} and, especially, {\it Chandra} 
high resolution X-ray images show clear evidence of  
AGN-gas interaction from giant elliptical galaxies (gE) to clusters
(\citealt{boe93,bla01,fij01,jon02}; \citealt{mcn07} and references therein).
Evidences of such activity have also been identified in many 
groups (\citealt{all06,mor06,jet08,gast09}).  
The fairly common presence of X-ray cavities, 
often coincident with lobes of radio emission
connected to the core of the central galaxy by a radio jet, indicates 
that AGN inject energy in the IGM 
in kinetic form (outflows) and as relativistic particles. 

Moreover, from dozens of giant ellipticals (gE) 
(\citealt{nul07}) to rich clusters (\citealt{raf06,raf08}) 
there is a continuous scaling relation
between the AGN power,
associated with the cavity formation, and the core X-ray luminosity.
This trend is more evident in low mass systems.
Although this connection
is likely to be a necessary but not a sufficient requirement for a
successful heating scenario, it strongly suggests
that the dominant heating process manifests itself
generating bubbles in the ICM, through some kind of `directional' energy input,
such as jets or collimated outflows (rather than spherically symmetric forms).

AGN outflows seem to be a very promising mechanism to simultaneously explain
the quenching of cooling flows (and star formation), the deviation of scaling relations
from self-similar laws and the presence of X-ray cavities and weak
shocks in the intergalactic (interstellar) medium. 
As anticipated before, the impact of this
type of heating is
much more evident in less bounded systems, such as groups.
We can therefore speculate that gentler outflows are more appropriate
to regulate the thermodynamic evolution of the IGM, 
while strong bursts should easily deform the main properties. \\

Regarding clusters, we can account for several variegated computations 
(e.g. \citealt{rus04,omm04,bru05,brm06,ste07,gas09,gas11} among others), 
but little work has been done to explore the role of AGN heating on the 
intragroup medium, especially from a theoretical and numerical point
of view. \citet{brm03} studied thermal heating (plus conduction) in 2D simulations,
showing that this type of feedback, while very efficient in stopping the cooling
process, often generates negative temperature gradients, with the total erasure of
galactic cooling flows commonly observed in gE (\citealt{sun05,sun07}).
Carrying out cosmological
simulations (\citealt{puc08,mcc10}) does not permit, nowadays, a detailed
analysis of the feedback dynamics in the inner cool core, mainly
because of the lack of resolution, and are thus more suited to study global
sample-averaged properties. Further, none of the above mentioned works has
investigated a purely mechanical feedback. 

In this paper we will therefore describe, for the first time,
3D self-regulated hydrodynamical simulations of
AGN outflows in an exemplary galaxy group, lasting at least 7 Gyr. 
The positive results of similar simulations
for rich galaxy clusters (G11) are a strong motivation for a detailed
analysis of this (mechanical) feedback mechanism in lower mass systems.

As in G11, we assume that AGN outflows are the key element in solving the cooling
flow problem. The physical foundation is that the relativistic jet, produced by the
active nucleus, entrains ambient gas, strongly decelerating within a
scale-length of just a few kpc.
Such subrelativistic AGN outflows have been observed in several
cD galaxies in every band, thanks to blueshifted absorption lines:
optical (\citealt{nes08,nes11}), UV and X-Ray
(see \citealt{cre03} for a review), 21-cm (\citealt{mor05,mor07};
see also the references in G11). They occur at a distance of a kpc with
velocities around $10^3-10^4$ km s$^{-1}$. The geometry of the bipolar 
outflows is still unclear, with usually narrow opening angles, and
possibly some kind of precession.

The timing of the feedback is also an important issue. From observations and from a
`thermostat' assumption the duty cycle has to be similar to the cooling time, around
$10^7$ yr. We will simplify the feedback self-regulation, testing hot (Bondi)
or cold accretion models with different mechanical efficiencies. Fully
3D simulations are 
required to compute the chaotic and turbulent flow, with associated
instabilities.


In summary, this work will focus on a set of heated cooling flow models
with a variety of AGN feedback mechanisms, aimed to answer the
following key question: {\it are AGN outflows
able to prevent the IGM from cooling
and at the same time preserve the observed cool core appearance for many Gyr?}

\section[]{The Computational Procedure}
In order to study the behaviour of AGN outflows in galaxy groups,
we carried out several simulations with a
substantially modified version of FLASH 3.2
(\citealt{fry00}), a popular 3D adaptive mesh refinement (AMR) code.
FLASH uses the Message-Passing Interface 
(MPI) library to achieve portability and efficient scalability on a
variety of different 
parallel high-performance computing systems. The simulations were usually
run on 128 processors of IBM P575 Power 6 (SP6) at CINECA supercomputing centre. 

We tested different numerical schemes to
solve the hydrodynamic Euler equations. As in G11, we widely used the 
Piecewise-Parabolic Method (PPM) 
solver (\citealt{cow84}), 
particularly appropriate to describe shock fronts. It uses directional
splitting
to advect quantities (\citealt{str68}).
This makes the instantaneous injection of the outflow not trivial, since we must
manage two equal timesteps with a sweep $x-y-z$ and then $z-y-x$. Thus,
we decided to also adopt other unsplit methods (single timestep evolution).
We tested Roe and Lax-Friedrichs, with different slope limiters (van Leer, minmod, etc.).
They work pretty well for smooth flows, but not for a powerful jet ignition, producing unstable
results also with low Courant-Friedrichs-Lewy (CFL) numbers ($< 0.1$). 
On the contrary, the HLLC solver (\citealt{tor99}) is very stable also
for higher CFL numbers ($\sim 0.5$), with
a good description of shocks and contact discontinuities. In the end, results with both
methods are comparable, with PPM being more accurate (3rd order), while
HLLC faster and more manageable. In many cases we preferred accuracy over speed.

The innovations implemented in FLASH reside, mainly, in
several source and sink terms added to the usual hydro-equations.
In conservative form:
\begin{equation}\label{cont}
\frac{\partial\rho}{\partial t} + \bmath{\nabla}\cdot\left(\rho \bmath{v}\right) = \alpha \rho_{\ast} -q\frac{\rho}{t_{\rm {cool}}} + S_{1,\rm{jet}}\:,
\end{equation}
\begin{equation}\label{mom}
\frac{\partial\rho \bmath{v}}{\partial t} + \bmath{\nabla}\cdot\left(\rho \bmath{v} \otimes \bmath{v}\right) + \bmath{\nabla}{P} = \rho\bmath{g}_{\rm {DM}} + S_{2,\rm{jet}}\:,
\end{equation}
\[
\frac{\partial\rho \varepsilon}{\partial t} + \bmath{\nabla}\cdot\left[\left(\rho \varepsilon + P\right) \bmath{v}\right] = \rho\bmath{v}\cdot\bmath{g}_{\rm DM} + \alpha \rho_{\ast}\left(\varepsilon_0 +\frac{\bmath{v}^2}{2}\right) \nonumber
\]
\begin{equation}\label{ene}
\quad\quad\quad\quad\quad\quad\quad\quad\quad\quad\quad -\: n_{\rm{e}} n_{\rm{i}} \Lambda(T,Z) + S_{3,\rm{jet}}\:,
\end{equation}
\begin{equation}\label{eos}
P = \left(\gamma -1\right)\rho \left(\varepsilon-\frac{\bmath{v}^2}{2}\right)
\end{equation}
where $\rho$ is the gas density, $\bmath{v}$ the velocity, $\varepsilon$ the specific 
total energy (internal and kinetic), $P$ the pressure, $\bmath{g}_{\rm
  {DM}}$ the gravitational acceleration (due to the dark matter halo
plus the central elliptical, see below), and 
$\gamma = 5/3$ the adiabatic index. 
The temperature is computed from $P$ and $\rho$ using (\ref{eos}), 
with an atomic weight $\mu\simeq 0.62$, appropriate for a totally
ionized plasma with 25\% He in mass.

In G11 we describe in-depth all the source and sink terms (right hand side of
the first three equations). Here we give a summary.

Radiative cooling is modeled according to \citet{sud93} cooling function $\Lambda(T, Z)$
for a fully ionized plasma ($n_{\rm {e}}$ and $n_{\rm {i}}$ are the
number density of electrons and ions). For typical groups, the central metallicity $Z$ is
about solar (e.g., \citealt{rap09}), with a negative gradient at larger radii.
For simplicity, we set the abundance equal to 1 Z$_\odot$.

Supernovae Ia (SNIa) and stellar winds (SW), in the central elliptical galaxy
are implemented
following \citet{brm02}, with a stellar mass loss rate $\alpha(t)$ 
dominated by stellar winds $\propto t^{-1.3}$.

The key prescription for treating the cold gas is the dropout term (Eq. (\ref{cont})),
proportional to $q=2 \exp (-(T/5\times10^5)^2)$.
It removes the gas below $T=10^{4-5}$ K, whose
physical evolution cannot be properly followed by our code\footnote{Tests 
without the dropout show the artificial assembly of cold clumps
near the center.}.
Because it has been proved that the total mass of cooled gas is insensitive to
the presence of the dropout term or its functional form (see
\citealt{brm00}), we can still calculate the gas cooling rate accurately.

Outflow source terms $S_{1,2,3,\rm{jet}}$ 
will be explained for every type of feedback in Section 2.2. Injection will be done
with two different methods:
directly into the domain (without a mass inflow) 
or through boundary condition at $z=0$ ($S_1 > 0$). 

We calculate the flow evolution for 7 Gyr.
Local cosmological systems have 
very different formation times and are
subject to relatively frequent mergers (\citealt{cow05}).
We conservatively decided to run our simulations for such a long time
to take into account the systems which do not undergo
major mergers since $z\sim 0.7$. Moreover, cool cores 
in massive objects
may survive major mergers (e.g., \citealt{bur08}), implying 
that the central AGN feedback is continuously needed. 
 
Note that several
feedback schemes could delay excessive gas cooling for only 1 Gyr
before failing (see G11). Thus, we stress that a few hundreds Myr 
evolution is not sufficient to
make predictions on the heating versus cooling regulation and 
might lead to misleading extrapolations.

\subsection[]{The group model and initial conditions}

\begin{table} \label{params}
\caption{Parameters and properties of the most relevant models.}
\begin{tabular}{@{}lcccccc}
\hline 
Model  & Feedback              &  efficiency ($\epsilon$)       & Notes \\ 
\hline
 CF    & no AGN heating        &      -           &   -               \\
 Bc1em2 & Bondi & $10^{-2}$               & continuous  \\    
 Bc5em2 & Bondi & $5\times10^{-2}$  & continuous  \\
 Bc1em1 & Bondi & $10^{-1}$               & continuous  \\
 Bi1em3 & Bondi & $10^{-3}$                & cold timing   \\
 Bi1em2 & Bondi & $10^{-2}$                & cold timing   \\  
 Bi5em2 & Bondi & $5\times10^{-2}$   & cold timing   \\      
 Bi1em1 & Bondi & $10^{-1}$                & cold timing   \\
 C5em5 & cold & $5\times10^{-5}$       & -    \\
 C1em4 & cold & $10^{-4}$                    & -    \\
 C5em4 & cold & $5\times10^{-4}$       & -    \\
 C1em3 & cold & $10^{-3} $                   & -    \\
 Int510l   & 5 - 10 Myr cycle &    -             & $P_{\rm j}=10^{-5}P_{\rm Edd}$ \\
 Int510m & 5 - 10 Myr cycle &    -             & $P_{\rm j}=10^{-4}P_{\rm Edd}$ \\
 Int510h  & 5 - 10 Myr cycle &    -             & $P_{\rm j}=10^{-3}P_{\rm Edd}$ \\
 Eth50    & thermal (+ Bc)  & $5\times10^{-3}$ & 50\% $E_{\rm th}$     \\
 IO1-9     & InOut (+ Bc)    & $10^{-3}$              &  $\dot{M}_{\rm in}=0.1 \dot{M}_{\rm acc}$ \\
 IOen40  & InOut (+ Bc)    & $5\times10^{-3}$  & $40\dot{M}_{\rm out}$ (IO3-7)    \\
 IOen80  & InOut (+ Bc)    & $5\times10^{-3}$  & $80\dot{M}_{\rm out}$ (IO3-7)     \\
\hline

\end{tabular}
\end{table}

We choose NGC 5044 as the template for a typical X-ray bright
group of galaxies:
all the results we present should be relevant for any object 
in this category.
NGC 5044 was one of the first cool core groups observed,
due to its high X-ray brightness, with low redshift 0.009
(see \citealt{buo03,gast09,dav09}, for recent
Chandra and XMM data). 
The estimated X-ray bolometric ($0.1-100$ keV) 
luminosity is  around $2\times10^{43}$ erg s$^{-1}$,
with a virial mass of $\sim 4\times10^{13}$ $\msun$.
The large scale morphology is very
smooth and nearly spherical, despite a little disturbance in the form of a south-eastern cold front.
At the contrary, the core ($\sim10$ kpc) has been strongly 
perturbed by recent outbursts from the central
AGN. The group presents many small radio quiet cavities with a nearly isotropic
distribution and moderate mechanical power ($\sim10^{42}$ erg s$^{-1}$).
The two biggest cavities at larger radii ($10-20$ kpc) are instead
filled with radio emission at 235 MHz (\citealt{dav09})
and the associated mechanical power seems to be able to 
balance radiative losses.
The NGC 5044 X-ray image also shows several cold filaments
coincident with H$\alpha$ and dust emission, indicating
a physical connection between the various gas phases (\citealt{gast09}).
The GMRT observation (610 MHz) reveals the presence
of extended radio emission with a torus-like morphology, threaded by the largest filament
(probably cold material being uplifted from the center).

The cooling time of the hot gas within the central 2 kpc is just $\sim4\times10^7$ yr.
In the absence of feedback, a pure cooling flow model predicts tens of $\msun$/yr
(see Eq. (1)), while observations suggest the cooling rate is at least an
order of magnitude lower, less than a few $\msun$/yr (\citealt{dav09}).
This is the key requirement for our computed models.

This galaxy group consists of a luminous giant elliptical galaxy (NGC 5044)
surrounded by a cluster of $\sim150$ low luminosity dwarf galaxies, mostly of early type. 
Thus, we choose to model the elliptical galaxy with a de Vaucouleurs profile (\citealt{mem87}) with
total stellar mass $M_{\ast}$ $\sim 3.4 \times 10^{11}$ 
M$_\odot$ and effective radius $r_{\rm {e}}$ $\sim 10$ kpc (\citealt{buo04}).\\

We start our calculations with the hot gas in spherical hydrostatic equilibrium 
in the potential well generated by the total system mass.
We use the observed $T(r)$ and $n(r)$ (\citealt{buo03,buo04}, see the dotted
lines of Fig. 1) to calculate 
the total gravitational potential under the assumption of hydrostatic equilibrium.
At the virial radius the gas fraction is $\sim 0.11$, a reasonable
value for X-ray bright groups (\citealt{mat05}).


The computational rectangular 3D box in all of our models extends slightly beyond the group
virial radius, $R_{104}$. We simulate the $z\ge 0$ half-space with symmetric boundary
condition at $z=0$, while elsewhere we set prolonged initial
conditions with only outflow allowed.
Despite the AMR capability of FLASH, we decided to use a number of concentric
fixed grids in cartesian coordinates. This ensures a proper resolution of 
the waves and cavities generated in the cluster core by the AGN outflows.
We use a set of 10 grid levels (basic blocks of $8\times8\times4$ points),
with the zone linear size doubling among adjacent levels.
The finest, inner grid 
has a resolution of 488 pc and covers a spherical region of $\sim20$ kpc in radius.
In general, grids of every level extend radially for about 40 cells.
This simple method is the best way found to cover large spatial scales
(hundreds pc up to Mpc) and at the same time integrating the system for several Gyr,
using moderate computing resources.

\subsection[]{Outflow generation}
We adopt a purely mechanical AGN feedback in form of nonrelativistic,
collimated outflows. The implementation is similar to the galaxy cluster
outflows of G11. 

In considering massive slow outflows, we are implicitly assuming that the relativistic
jet entrains some ICM mass ($M_{\rm act}$), or in a similar way,
that radio jets are highly relativistic on pc scale, 
but rapidly decrease to subrelativistic velocities within few kpc from the black hole
(\citealt{gio04}).

As in G11 we show results only for models
with cylindrical jets, with velocity
parallel to the $z-$axis. We have calculated few simulations with
conical (or precessing) outflows 
and verified they have a similar
impact on the global properties of the flow. In fact, the pressure of
the surrounding gas collimates the outflows within few tens kpc.

In this paper we will focus mainly on the two most successful models
adopted for clusters in G11: cold and Bondi feedback.

In the cold feedback\footnote{See also
\citealt{sok06} for a similar type of heating.}
an outflow is activated only when gas cools to very low
temperature within a 
spherical region $r<3$ kpc, and drops out
from the flow. If $\Delta M_{\rm cool}$ is the mass cooled in the
aforementioned region during a single timestep, 
the injected kinetic energy (in the following timestep) is given by:
\begin{equation}\label{cold}
\Delta E_{\rm jet} = \epsilon\,\Delta M_{\rm cool}\, c^2.
\end{equation}

This energy 
is given to the hot gas located in a small region at the centre of the grid
(the `active jet region'), whose size is always\footnote{A
slightly smaller or bigger jet does not alter the global evolution,
see G11.} 
$2\times4$ cells ($\sim$1 kpc wide, 2 kpc high),
containing a hot gas mass $M_{\rm act}$: 
\begin{equation}\label{cold2}
\frac{1}{2} M_{\rm act} v_{\rm jet}^2 = \Delta E_{\rm jet}.
\end{equation}
  
We will see
that the frequency and strength of the feedback events
strongly depend on the mechanical efficiency $\epsilon$, which has typical values
$10^{-4} - 10^{-3}$ (see Table 1), almost an order lower than galaxy clusters.

In the Bondi feedback the outflows are triggered by 
the usual prescription: 
\begin{equation}\label{Bondi}
\dot M_{\rm B} = 4\pi (GM_{\rm  BH})^2\rho_0/c^3_{{\rm s}0},
\end{equation} 
where $\rho_0$
is the volume-weighted hot gas density calculated within $\sim 2.5$ kpc, 
while $c_{{\rm s}0}$ is the mass-weighted sound speed in the same region.
The assumed SMBH has a mass of $3\times10^9$ $\msun$. The outflow
energy
is then calculated with eq. (7), using $\dot M_{\rm B}$ instead of
$\Delta M_{\rm cool}$.
Needless to say, the Bondi radius, $r_{\rm B}\sim 80$ pc, is smaller
than our resolution limit, so we refrain to attach a strict
physical meaning  to $\dot M_{\rm B}$. 
In this sense the high mechanical efficiencies adopted for
Bondi models are due to the fact that the accretion should be considerably larger (because
of higher inner $\rho_0$ and lower $c_{{\rm s}0}$).

Besides the two main feedback schemes, we experimented some variations
in the AGN generation, such as fixed intermittency, thermal feedback,
direct linking between outflowing and accreting mass.
These special models are often not successful and they will be briefly
explained through Sections 3.4 - 3.6
(see also Table 1 for a summary).\\

For most of the presented methods we applied injection directly in the domain with
$M_{\rm act}$, as stated above. In other simulations
we tested also the injection of mass, momentum and energy
flux through a `nozzle' in the grid boundary at $z=0$. 
This way, the jet power is expressed as:
\begin{equation}\label{nozzle}
\frac{1}{2}\, (\rho_{\rm jet} v_{\rm jet} A_{\rm n})\, v_{\rm jet}^2 = \epsilon\,  \dot{M}\, c^2,
\end{equation} 
with a nozzle area $A_{\rm n}$ 
of $2\times2$ cells and $\rho_{\rm jet}\sim 0.1$ the initial central gas density.
We fix the temperature
of the jet to very low values, compared to the IGM, in order to keep
injected thermal energy on negligible levels compared to
the kinetic flux.

As noted also in G11, the results do not greatly depend on the method
of injection. However, the physical approach is a bit different:
in the $M_{\rm act}$ method we are directly modelling the entrainment, 
while in the nozzle injection we are
specifying the properties of the `subgrid jet'.

\section[]{Results}

In this Section we report the results of the various simulations, exploring
different types of feedback and parameters.
We have analysed in detail
the long term behaviour
of models similar to the most successful ones in G11. 
Some of the properties of unsuccessful - but pedagogical - models
are covered in G11. 

We stress again that the main objective of the present work is investigating the global
properties of the flow, such as cooling rates and azimuthally averaged profiles
(density, temperature, etc.). Cavities, shocks and iron abundances will be
studied in-depth elsewhere through a dedicated  set of simulations at higher
resolution.

\subsection[]{Pure cooling flow}

As a fiducial model, we ran a simulation without AGN feedback, i.e. a pure
cooling flow (CF). The results are shown in Fig. 1.

\begin{figure}
\centering
\includegraphics[width=55mm]{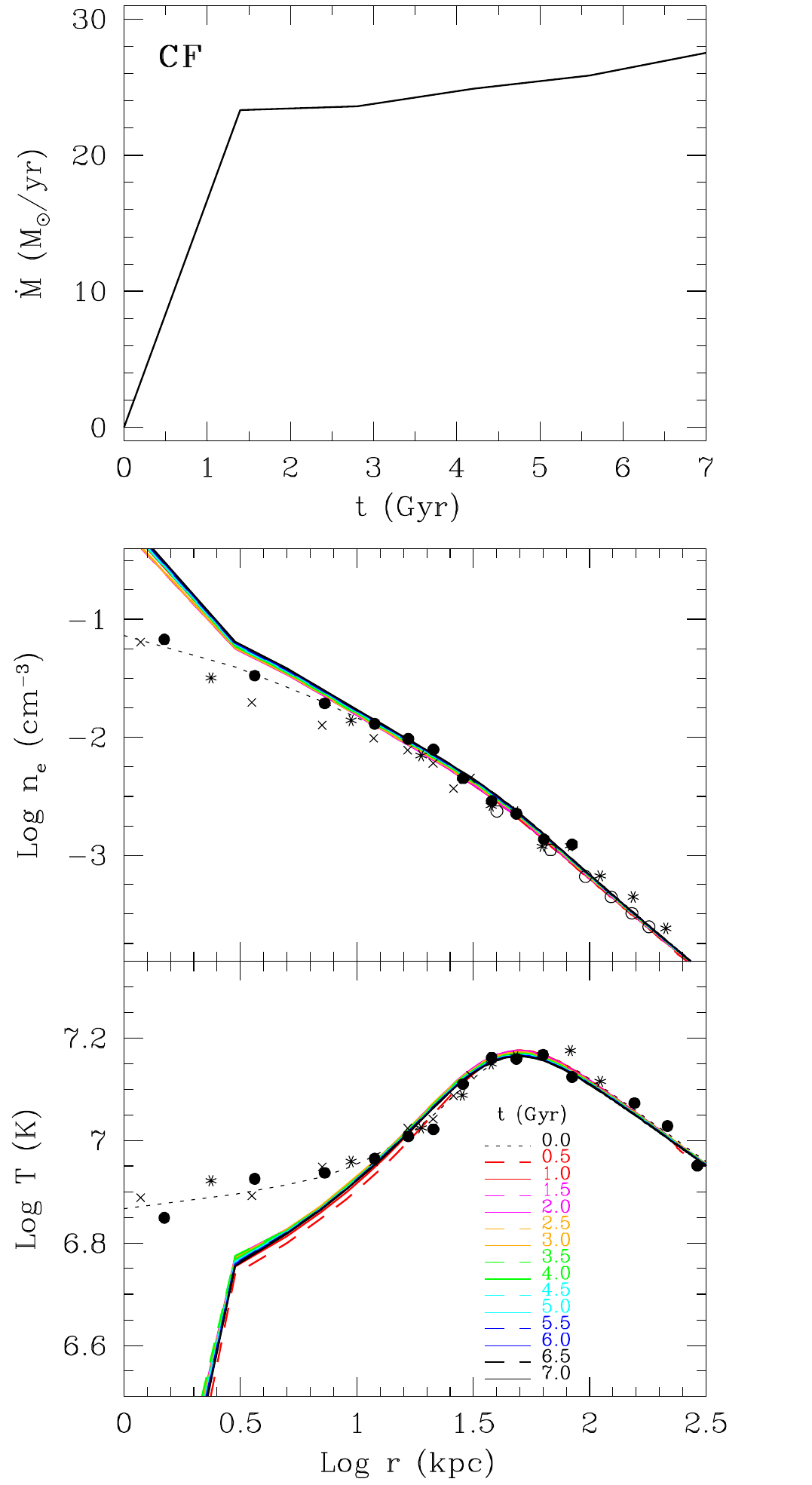}
\caption{Evolution of CF model (no AGN feedback). Top panel: 
gas cooling rate versus time.  Middle and bottom panels:
temporal evolution of the gas (electron) number density and 
mass weighted temperature profiles,
respectively. The profiles are displayed at 15 different times, as indicated
in the lowermost panel.  {\it Chandra} and {\it XMM-Newton} 
observational data of NGC 5044 are represented by
star and cross points (\citealt{buo04}), 
filled circles (\citealt{buo03}); open circles
indicate {\it ROSAT} data (\citealt{dav94}).} \label{fig:cf} 
\end{figure}

The flaws of the classical cooling flow model are
clearly depicted in this evolution. Both the density
and (mass weighted) temperature profiles steepen considerably in 
the central region: the radiative losses induce a subsonic inflow of
the gas, increasing the emissivity in a vicious cycle. This trend
is relatively more evident and fast in the galaxy group, rather than
in the cluster. After just 1 Gyr the cooling rate has already approached
a quasi steady state, with an asymptotic value $\sim 25$ $\msun$/yr.
This is the clearest discrepancy with observational data and
one of the main failure of the standard cooling flow scenario.

The other two main observables, $T(r)$ and $n_{\rm e}(r)$, are
also clearly wrong in the inner $\sim15$ kpc (see Sec. 1 and 2.1 for
the observational references). Very cold gas, down
to $10^5$ K (which is soon dropped in our simulation), concentrate
in the nucleus and generate non consistent gradients.
Our feedback models have to reduce this steepening, but preserving
the cool core appearance at the same time. As said in Sec. 1, galaxy groups present
a more variable inner temperature profile respect to clusters.
Nevertheless, the heating must not destroy the inner structure
(see also \citealt{brm02,brm03}),
a quite demanding requirement.

A typical signature of such a strong cooling flow is also the temporal increase
of the bolometric X-ray luminosity: from $\sim 1.9\times10^{43}$ erg s$^{-1}$
up to $\sim 2.6\times10^{43}$ erg s$^{-1}$ at 7 Gyr.

It is interesting
to investigate the global energetic budget, because the common
(misleading) sense would probably associate the irradiated energy
with, mainly, the drop in internal energy.
However, the internal energy within
$r_{\rm vir}$ decreases by $\sim 7 \times 10^{58}$ erg (that is, by
less of $1$\%), while the
potential energy drops by $\sim 6 \times 10^{60}$ erg, considering both the
hot gas remaining in the grid
and the cooled gas at the centre of the cluster.
The kinetic energy
stays always around $\sim 10^{57}$ erg and is therefore negligible.
The conclusion is that energy is radiated away ($E_{\rm rad}\sim 6 \times 10^{60}$ 
erg in 7 Gyr) mainly at the expense of the potential energy of the IGM. This behaviour was less
evident in the cluster, but still present.

In order to test the effect of numerical resolution on the CF results,
we calculated a CF simulation with a resolution
three times higher: the results are almost identical to the
lower resolution run described above. Therefore, with the adopted
resolution our models are clearly in the convergence limit.\\

After placing the frame of the requirements for a successful feedback model,
in the following we will describe the simulations
with heating linked to AGN outflows.
In the next Section, we will begin with the two best Bondi models and then we will discuss,
from higher to lower efficiencies, computations with worser results,
alternating continuous and intermittent type of feedback.

\subsection[]{Bondi feedback (or entropy-regulated)}

\begin{figure*}
\includegraphics[width=0.698\textwidth]{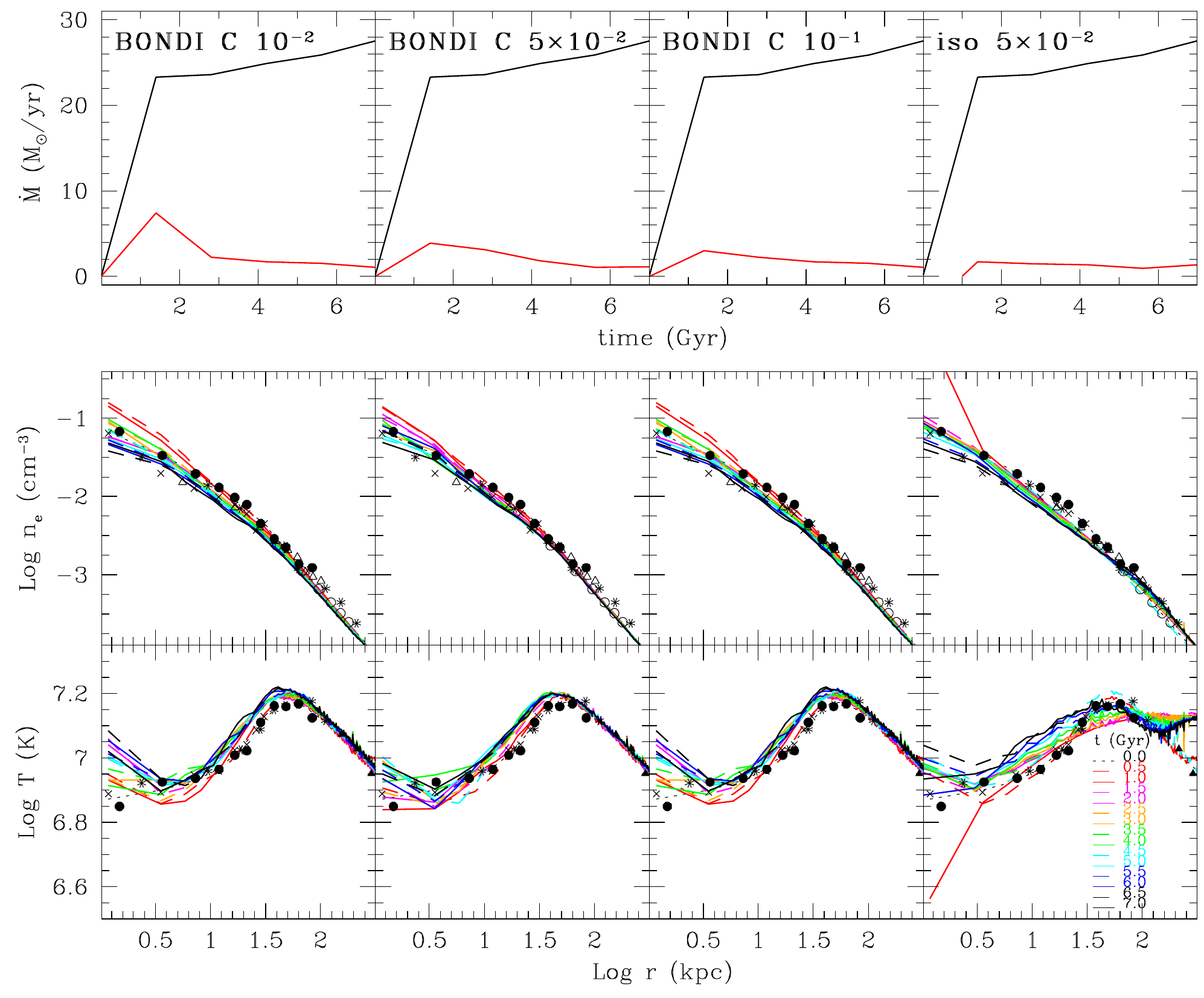}
\includegraphics[width=0.699\textwidth]{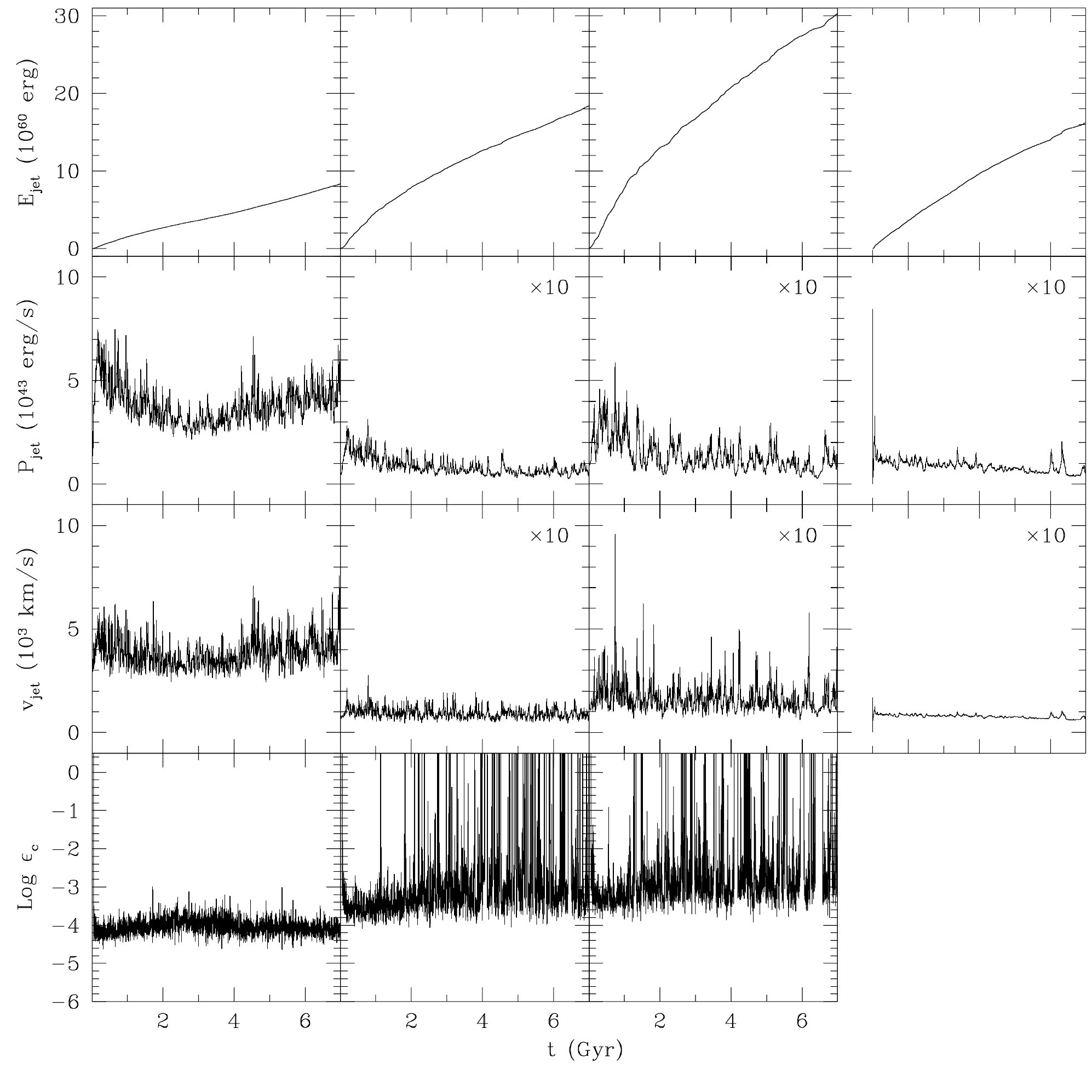}
\caption{Evolution of Bondi (continuous) feedback models with
different efficiencies (the latter being the isothermal start).
Panels in the first three rows (top to bottom)
represent the same quantities as in 
Figure 1. Panels in the other four rows show:
total injected kinetic energy (in the
half-space $z>0$),
outflow power and velocity, associated cold efficiency (see text). The
label `$\times 10$' means that the units in the ordinate axis have been
multiplied by 10.} \label{fig:Bondi_cont}
\end{figure*}

\subsubsection[]{Model Bc5em2, $\epsilon_{\rm B} = 5\times10^{-2}$}

We begin to analyse the results of the simulations with
AGN feedback turned on. 

The Bondi accretion theory (\citealt{bon52}), although highly idealized
with respect to the complexity expected in real accretion systems,
has been widely used to estimate accretion rates on supermassive
black holes, adopting  Eq. (\ref{Bondi}).

Despite our resolution is far from capturing the
effective Bondi radius (tens of pc) and despite the fact that the
radiative cooling is important in the central region,
we think that Bondi accretion is still a
useful, simple feedback prescription. In fact, the accretion rate is simply proportional to
$K^{-3/2}$, where $K = T/\rho^{2/3}$ is the 
entropy parameter. Thus, Bondi regulation is intrinsically
a good `thermostat': when the gas is cold and dense, the feedback is stronger, and vice versa.

The first good model that we obtained is Bondi accretion with efficiency 
$\epsilon = 5\times10^{-2}$
(Fig. \ref{fig:Bondi_cont}, second column), using the `entrainment' injection in the active region.
The zone for the weighted quantities is $\sim2.5$ kpc. 
The overall feedback is gentle and moderate, exactly
as speculated. 

The mean cooling rate is well below $10 \%$ of the pure cooling flow,
with only a transient peak ($\sim3.5$ M$_\odot$ yr$^{-1}$) at 1.5 Gyr.
An important feature for a successful continuous feedback, as
seen in G11, is stability in time. The moderate
power stays around $10^{44}$ erg s$^{-1}$, without strong variations,
which allows to balance the instantaneous cooling rate in an efficient way.
 
The mass-weighted temperature
profiles are in agreement with observations 
in the nucleus of the group, with a smooth and 
almost flat curve. 
In the outer regions tiny acoustic oscillations are present,
while between 10-30 kpc
the continuous injection of energy tends to slightly overheat this
region, especially at late times.
The density profiles also show little variation from the initial 
observed ones, an indication that the gas is not accumulating and 
peaking in the core.

Another relevant consideration is that the Bondi rate is
always sub-Eddington (between $10^{-3}-10^{-2}$ Eddington rate). 
Observations
of X-ray bright elliptical galaxies (e.g. \citealt{all06})
also point towards similar values.

The final injected energy is relatively low, $\sim1.8\times10^{61}$ erg
($3.6\times10^{61}$ erg for the bipolar outflow),
compared to the total `available'
BH energy $1.8\times10^{62}$ erg; 
roughly estimated as $E_{\rm BH}\sim 0.1\,M_{\rm BH}\,c^2$, with
$M_{\rm BH}=10^9$. This will be a typical feature of galaxy groups, in
contrast to the high input energies of massive clusters, whose models 
can sometimes exceed the above quantity. 

Furthermore, we tested several initial conditions and we concluded that
the dynamics and results of our feedback models are,
overall, not distorted by relatively
different initial temperature and gravity profiles. For example, in
Fig. \ref{fig:Bondi_cont} (fourth column), we started the computation with
an isothermal profile ($\sim1.35\times10^7$ K), retrieving the density
structure from a theoretical NFW plus deVaucouleurs potential. After the
first Gyr of pure CF, the Bondi feedback ($\epsilon_{\rm B} = 5\times10^{-2}$) 
is again very effective
in quenching the cooling flow and at the same time preserving the cool 
core appearance.

A possible riddle for all continuous models remains the absence of frequent jet-inflated
spherical cavities. The continuous AGN activity (mean $v\sim 10000$ km s$^{-1}$) carves a narrow tunnel of about 30 kpc in length,
although its density contrast with the environment is large only for $z\lta 10$ kpc (and not particularly evident in the surface brightness maps; see Sec. 4). 

\subsubsection[]{Model Bi5em2, $\epsilon_{\rm B} = 5\times10^{-2}$}

In a self-regulated (subgrid) mechanism 
we have to face substantially two problems.
The first one is how much power to link to the feedback, while
the second one is the right timing, i.e. when to turn on and off
the heating machine.

\begin{figure*}
\includegraphics[width=0.698\textwidth]{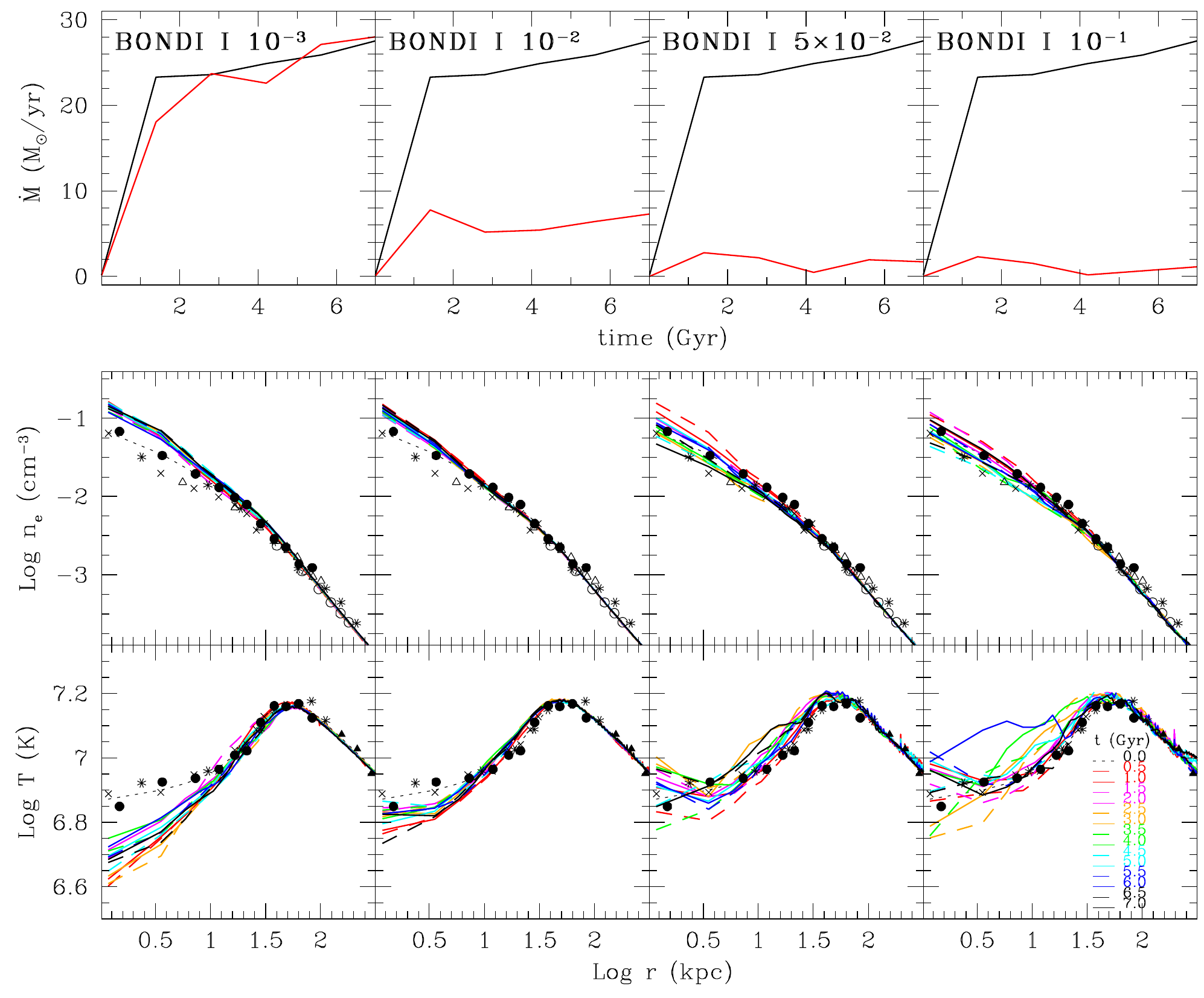}
\includegraphics[width=0.699\textwidth]{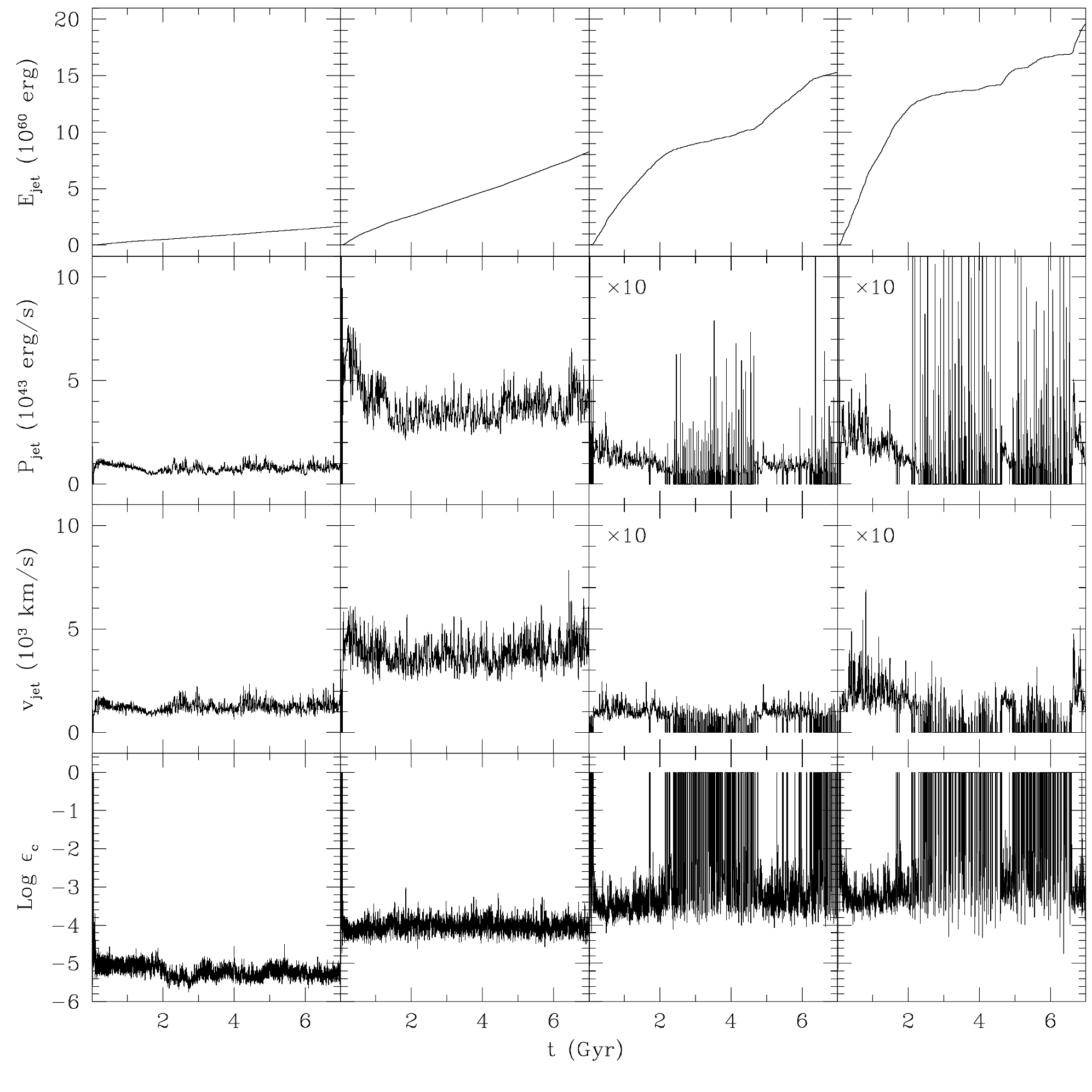}
\caption{Evolution of Bondi (intermittent) models with increasing efficiencies,
from left to right. The description of the plots are same as for Fig. 2.} \label{fig:Bondi}
\end{figure*}

In the following simulation (Figure \ref{fig:Bondi}, third column) we
link the activation of the AGN heating to the gas cooling, but use
the standard Bondi rate for the gas accretion onto the black hole.
In other words, an outflow is generated when $\Delta M_{\rm cold}\neq
0$, with an energy linked to $\dot M_{\rm B}$ in the usual way.

This feedback scheme tries to reproduce a self-regulated
intermittency of the outflows (ideal Bondi models can only be 
continuous),
with cycles of inactivity
corresponding to moments of high hot gas entropy. In fact, 
when the gas begins to cool, it will usually start
to flow towards the central black hole.

This way,
the temperature profiles are a bit more consistent with
the observational data, especially at intermediate radii.
Although the jets are intermittent, their frequency is high,
a condition that
grants the quenching of cooling rate ($\dot M_{\rm cool}\sim
0.5-2$ M$_\odot$ yr$^{-1}$). The density profiles do not
depart from the cool core status, without a strong steepening
(central values $\sim 0.1$ cm$^{-3}$).

Bondi accretion rates are almost identical to Bc5em2, with the total
energy injected in 7 Gyr being a bit less: $\sim1.5\times10^{61}$ erg.
The average jet power is also a factor of $\sim1.3$ less,
$\sim5.5 \times 10^{43}$ erg s$^{-1}$.

This model clearly solves the problem of jet-inflated
cavities in the IGM. The ouflows duty
cycle\footnote{The common duty cycle 
retrieved by observations can be quite
different from our numerical estimates, because the definition of an `active' AGN
highly depends on some luminosity threshold (in the X-ray, radio, etc.) and on
the properties of the sample.}
is roughly 80-85 per cent (in the whole computation). 
The pattern is not
very regular, in fact the cycle is more frenetic 
between 2 and 5 Gyr, when the central gas has been
heated efficiently.
Every time the jet (with velocities usually
of 8000-10000 km s$^{-1}$) turns on again,
a bubble of small size ($10-15$ kpc) is seen in the density map. 
Note that after
a period of quiescence the subsequent injection produces
a peak in the power, promoting the above mentioned cavity inflation 
through shock heating. The side effect consists in more evident
perturbations in some $T$ profiles (e.g. 3 and 7 Gyr), however
still compatible with typical galaxy group observations.

Finally, during runtime we calculated, at every timestep, the efficiency ($\epsilon_{\rm c}$) that
would have the cold feedback mechanism (bottom panel). 
The result is striking: on average,
the two schemes are almost coincident with $\epsilon_{\rm c}$ around $10^{-3}$,
except for few short timesteps, in which the cold feedback requires
a 100 per cent efficiency (because the cooled mass is almost zero). 

\subsubsection[]{Model Bc1em1, $\epsilon_{\rm B} = 10^{-1}$}

We begin now to describe models from higher to lower efficiencies, alternating
again continuous and intermittent Bondi methods.

The following simulation is very computationally expensive, because
the (continuous) outflow velocity rises often over $20000$ km
s$^{-1}$. It is shown in Figure 2, third column.
The cooling rate is quenched to fraction of $\msun$ yr$^{-1}$ 
after just a few Gyr.
The total injected
energy is $3\times10^{61}$ erg (about 1.5 times greater the
intermittent counterpart, next model Bi1em1). 
The consequence of this high energy deposition
is a progressive steepening in the temperature gradient, with a moderate
spike in the inner 4 kpc, where the jet is always active.
During some timesteps the outflow reaches $5\times10^{44}$ erg s$^{-1}$,
a quite strong power for Bondi accretion.

The compared efficiency analysis (bottom panel) shows that a cold feedback
triggering could not even be possible in 20-30 per cent of the evolution,
because it would break the unity threshold. In the other periods it would be
associated with a value around $2.5\times10^{-3}$.

\subsubsection[]{Model Bi1em1, $\epsilon_{\rm B} = 10^{-1}$}

The following (Fig. \ref{fig:Bondi}, fourth column) 
is another computationally expensive model, which
touches the limit of acceptability in the parameter space,
for Bondi models. The timing is again associated with the cold gas.
The cooling rate is very similar to the precedent continuous model (Bc1em1),
with the reasonable range of
0.5-1.5 $\msun$ yr$^{-1}$ range, declining at late times.
The density profiles are optimal, not altered
from the initial data. Temperatures start to feel the strong outbursts,
which have sometime power in excess of $10^{45}$  erg s$^{-1}$ (with 
velocity larger than $2\times
10^{4}$ km s$^{-1}$). In fact at later times the central gas is heated
up to $10^{7}$ K, during some event.

The associated cold efficiency has a common trend around $2.5\times10^{-3}$,
but with many oscillations after 3 Gyr, reaching easily unity.
Notice that these moments last very few timesteps, after which
$\epsilon_{\rm c}$ reset to lower values (in contrast to a first
impression of the tight plot).

\subsubsection[]{Model Bc1em2 and Bi1em2, $\epsilon_{\rm B} = 10^{-2}$}

When the efficiency is lowered to $\epsilon = 10^{-2}$
the heating provided by the feedback is considerable ($\sim
8\times10^{60}$ erg). This decreases
the cooling rate down to at least 30-40 per cent of that of the pure
CF (Figure \ref{fig:Bondi_cont}, first column).
The profiles are close to
the observed data, thanks to an always continuous outflow
with velocity $4-5\times10^{3}$ km s$^{-1}$ and power $4-5\times10^{43}$ erg s$^{-1}$.

The intermittent simulation is very similar to the continuous evolution (Fig. \ref{fig:Bondi},
second column),
except that, initially, the system is permitted to cool. Without this early opposition,
the outflows - with same power - can not halt the cooling so efficiently like
in the previous case. The feedback finds an equilibrium only at 4 Gyr, while at later
times it will slowly increase, on the contrary of a 
reasonable self-regulated model. 

The associated efficiency of a $\Delta M_{\rm cold}$ model
is near $10^{-4}$. In fact, we will see that the linked cold
feedback model will produce an almost identical outcome, i.e. weak heating
(Fig. \ref{fig:DMc}, second column).

\subsubsection[]{Model Bi1em3 (or Bc1em3), $\epsilon_{\rm B} = 10^{-3}$}

As a template for a typical weak model, a simulation
with Bondi efficiency of $10^{-3}$ produce a cooling
rate, which is identical to a pure CF run: $\sim25-26$
$\msun$ yr$^{-1}$ (Fig. \ref{fig:Bondi}, first column). 

The only clue for the presence of AGN heating is that
the profiles do not suffer a heavy decline with time, even if
they posses a remarkable steep gradient 
($T$ down to $4\times10^{6}$ K).

The outflow is always continuous (the model could be also named Bc1em3),
with a very low velocity
around 1000 km s$^{-1}$ and a power often under
$10^{43}$ erg s$^{-1}$. There is no possibility that
this model will become intermittent at any time, due to 
the constant central cooling.

\subsection[]{Cold Feedback ($\bmath{\Delta M_{\rm cool}}$ regulated)}

\subsubsection[]{Model C5em5, $\epsilon_{\rm c} = 5 \times 10^{-5}$}

The results presented in the previous Section for the Bondi
accretion models suggest that,  
compared to galaxy clusters, groups require lower
efficiencies of at least a factor of 10. Therefore, we start illustrating the
cold feedback mechanism for the
lowest adopted $\epsilon_{\rm c}$ and then moving to more efficient outflows.

Here we still use the entrainment injection method, with 
width and length of the active region 1 and 2 kpc (2 and 4 grid points), respectively.
In G11, in fact, we concluded that the size of the jet and the type
of injection do not change drastically the evolution of the feedback. 

\begin{figure*}
\includegraphics[width=0.698\textwidth]{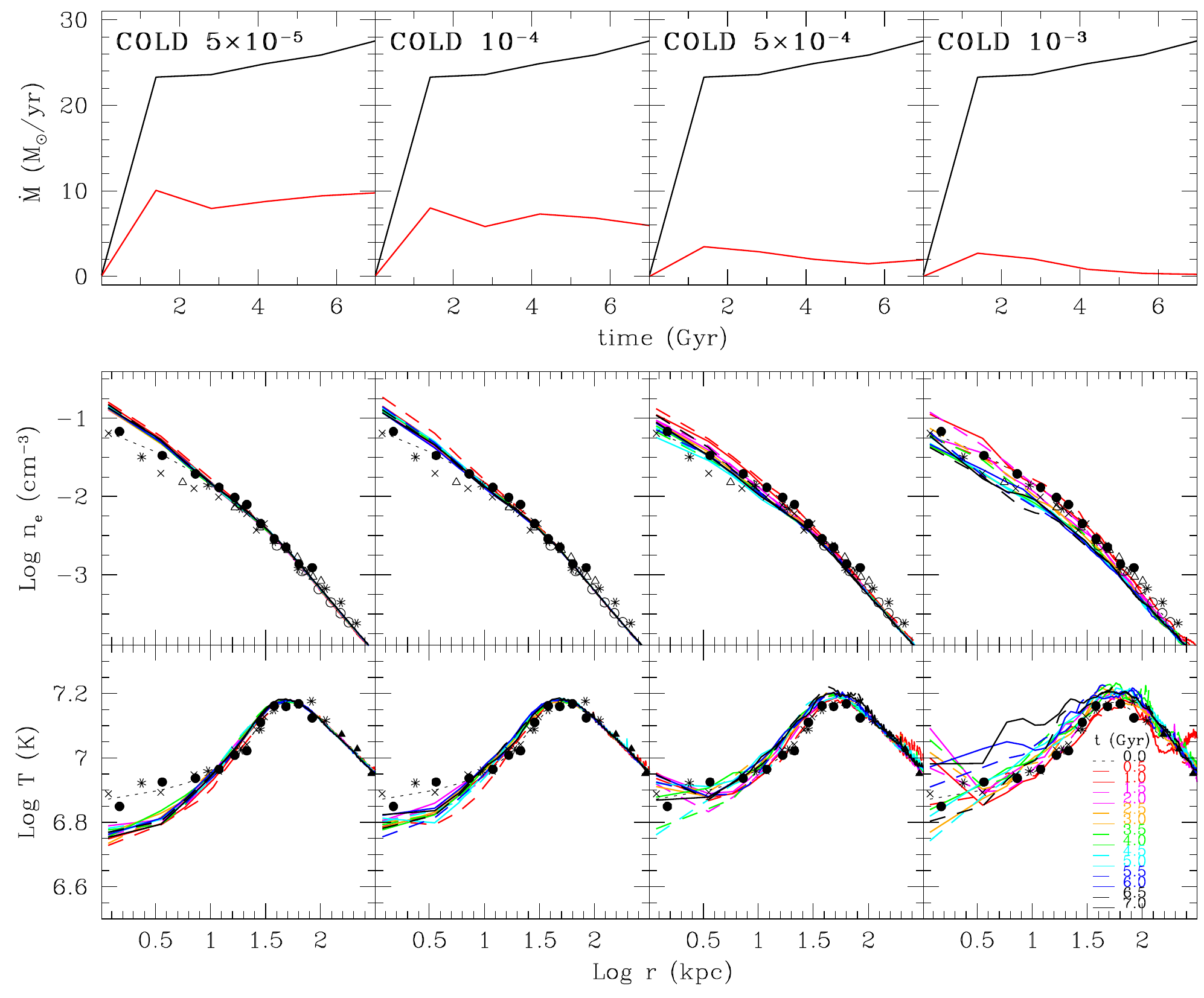}
\includegraphics[width=0.699\textwidth]{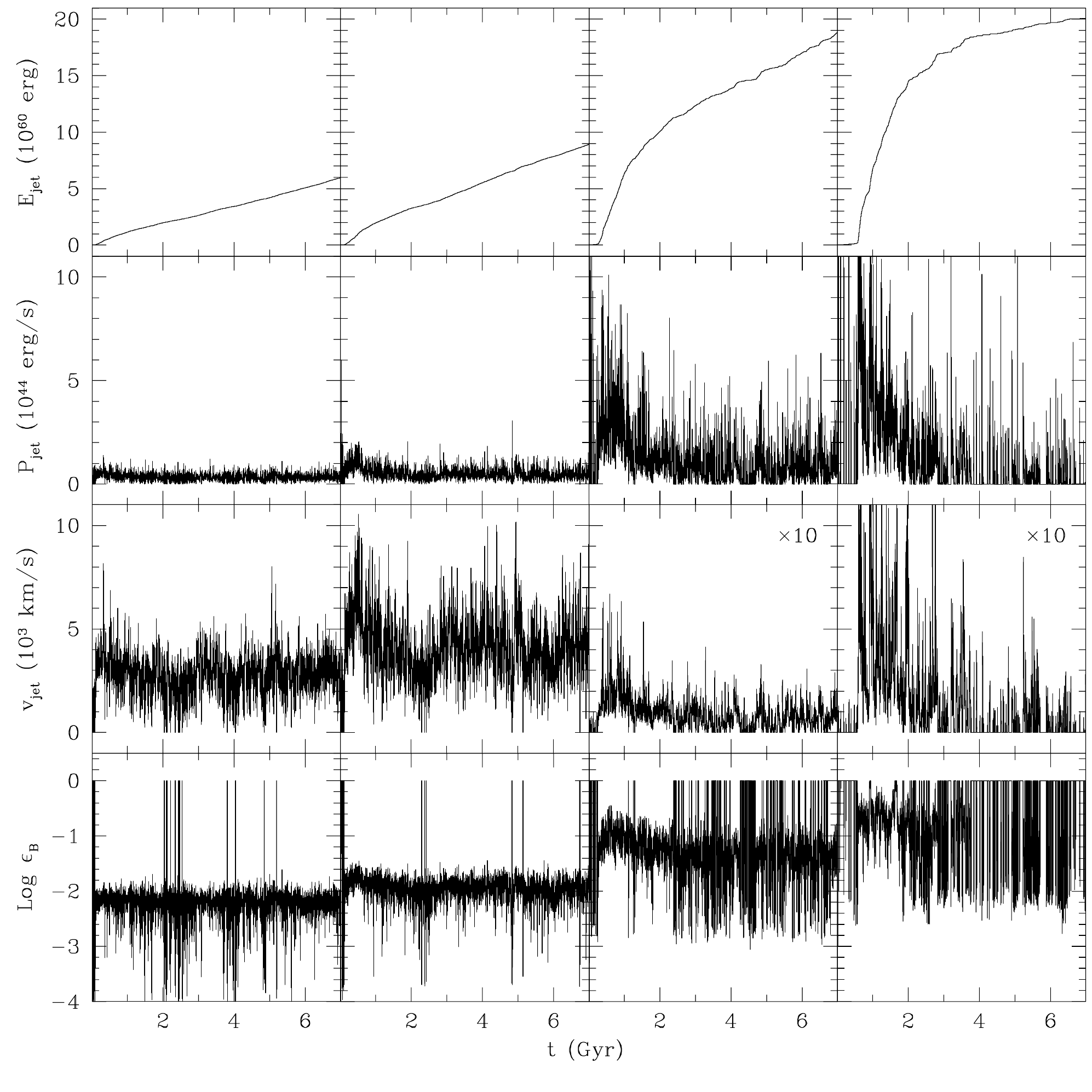}
\caption{Evolution of all cold accretion models with increasing efficiencies,
from left to right. The description of the plots are same as for Fig. 2.} \label{fig:DMc}
\end{figure*}

As shown in the first top-left panel of Figure \ref{fig:DMc}, 
the cooling rate settles
on a steady value around  $\sim 10$ $\msun$ yr$^{-1}$. This rate
is obviously unacceptable, although reduced to the 
$\sim 50 $\% of the
pure CF model value. With such a low efficiency, 
the azimuthally averaged, mass weighted temperature and density profiles
are only slightly modified by the heating and do present the excessive
accumulation of gas near the center, typical of the CF model.
In the
center the temperature is $\sim 5.6\times10^6$ K and the numerical densities
$\sim 0.18$ cm$^{-3}$.

The outflows,
typically have a power of $4-5\times10^{43}$ erg s$^{-1}$
and a velocity of $3000-4000$ km s$^{-1}$.
The total injected mechanical energy is in fact quite low, $6\times10^{60}$ erg,
and the outflows are almost continuous, because they are never
able to stop the cooling flow. Given the rare period of inactivity, the `duty cycle'
is $\sim 95$ per cent. 

At the end of the simulation
$6 \times 10^{10}$ M$_\odot$ have cooled and dropped out of the hot phase.
The vast majority of gas cools at the very center.
If all the cooled gas were accreted onto the central black hole,
as we have simplistically assumed in this scheme, the final BH mass would
result about an order of magnitude greater than the expected one.
In principle, we could avoid the problem of the excessive black hole mass
by assuming that only a fraction of the cooled gas actually accretes on it with
a higher efficiency (see Sec. 3.6, InOut models).

\subsubsection[]{Model C1em4, $\epsilon_{\rm c} = 10^{-4}$}

As anticipated, the next step is to increase the efficiency, in order
to find a more suitable heating power. 

In the following model, where the efficiency increased by a factor of
five, $\epsilon_{\rm c} = 10^{-4}$
we obtain in fact a jet power of $\approx 10^{44}$ erg s$^{-1}$ and
velocity oscillating around 5000 km s$^{-1}$ (see Fig. \ref{fig:DMc}, second column).
However, the cooling rate is only slightly reduced to $6-7$ $\msun$ yr$^{-1}$
and radial profiles tend to be similar to the ones of previous
model. The difference is that now the curves decline at 6 kpc, instead
of 10 kpc. Evidently, the outflows deposit their energy in more
inner zones.

We conclude that, even if the total injected energy reaches almost 
$10^{61}$ erg, at this stage the cooling flow is not quenched in a
successful manner. This is also underlined again by the almost continuous
presence of the outflows.

\subsubsection[]{Model C5em4, $\epsilon_{\rm c} = 5 \times 10^{-4}$}

At $5\times10^{-4}$ we finally recover a successful heating evolution.
As in previous computations, at early time (first Gyr) the
radiative cooling prevails over the outflow mechanical heating and
the cooling rate
reaches a peak of 4 $\msun$ yr$^{-1}$.
However the long term evolution presents different intriguing results.
After 2 Gyr the system becomes much more turbulent and chaotic,
inducing a reasonable amount of
mixing, that permits the gas circulation and reheating
in a very efficient way, also in the 5 kpc nucleus.

As shown in the third column of Fig. \ref{fig:DMc}, the temperature
in that region is almost constant, with values close to that observed for
NGC 5044 ($\approx 8\times10^{6}$ K; \citealt{buo03,dav09}).
This very important feature is also described by
the density curves: a gradual flattening of the central profile is seen through
the entire evolution. It is striking that the 7 Gyr (black) profile is practically superposed
to the initial fit. At that time (redshift = 0) the cooling rate is
$\sim 2$ $\msun$ yr$^{-1}$, about 7\% of the pure CF run, a significant suppression
which brings the value of $\dot M_{\rm cool}$ in reasonable agreement
with the observations
(\citealt{dav09,buo03,tam03}).

The shock waves generated by the outflows, with power usually between
$\sim 2-5\times 10^{44}$ erg s$^{-1}$, are not strong
enough to significantly alter the global
positive temperature gradient, although
small amplitude ripples are present.
These waves correspond to weak shocks associated with the jet propagation
and are visible up to a distance of $\sim 300$ kpc. They are not so evident
as in galaxy clusters because the energy involved here are two or three
orders of magnitude lower. In fact the total jet energy is
$1.8\times10^{61}$ erg. The gas is typically ejected with $v_{\rm jet}
\sim 10^4$ km s$^{-1}$.

Furthermore, an important difference between clusters and groups is 
the duration of each event. 
In cluster models, a duty cycle of 0.1 was quite usual,
because the very short super-Eddington bursts could stop instantly the 
cooling flow, with the consequence of perturbed profiles, near the ignition time.
Here, our estimated duty cycle is roughly 0.85. Therefore the situation is reversed:
successful cold feedback models in groups require an almost continuous heating, with
short pauses of tens Myr.

Notice that the Eddington luminosity,
$L_{\rm Edd}\approx 1.5 \times 10^{38}(M_{\rm BH}/M_\odot) \sim
10^{47}$ erg (for a $10^9$ M$_\odot$ black hole),
is far above the regime of the current model. In fact the
accretion rate oscillates between $10^{-3}-10^{-2} $ the Eddington rate.
As seen in Section 3.4, a $L_{\rm Edd}$ jet can easily erase the cool core
structure of the group. 

In summary, the analysed behaviour of C5em4 seems
to be similar to that the quiet Bondi feedback.
In the bottom panel we have calculated,
at every timestep, the associated Bondi accretion rate and the required
$\epsilon_{\rm B}$ to reproduce the same instant mechanical energy.
As a striking result, a Bondi efficiency of $\sim 5\times10^{-2}$, especially after
2 Gyr, is retrieved from this analysis. Only in a few events, the cold mechanism
detaches from the regular regime of Bondi, because of its intrinsic impulsivity.
Nevertheless the similitude is quite evident and indeed, if we compare the
two outcomes of both models (especially intermittent Bondi, Fig. \ref{Bondi},
third column),
they appear deeply connected also in a quantitative
manner.

\subsubsection[]{Model C1em3, $\epsilon_{\rm c} = 10^{-3}$}

Increasing the efficiency by a factor of 2
generates another positive model, with some features
approaching the borderline of a violent disruptive heating.

In fact, with a jet power up to $10^{45}$ erg s$^{-1}$,
the cooling flow is perfectly stifled, asymptotically decreasing
to a fraction of $\msun$ yr$^{-1}$ after a few Gyr
(Fig. \ref{fig:DMc}, fourth column). 

The beginning of the violent regime is quite evident in
the averaged profiles, which now tend to oscillate in a large
$\sim 50$ kpc region.
Overall, the cool core is conserved,
but the temperature indicates some overheating in the nucleus
at later times.

With our adopted active region, in a few outbursts, the velocity 
is greater than $10^{5}$ km s$^{-1}$ line, while the outflow mass rate is
several tens $\msun$ yr$^{-1}$, as in all previous $\Delta M_{c}$ models. 
This is a clear indication that
at higher efficiencies we will approach a relativistic regime and
a catastrophic heating (we will not show here results for 
$\epsilon_{\rm c}> 10^{-3}$ but see
fig. 5 of G11 for an analogous model for clusters). 
The equivalent Bondi efficiency for model C1em3 is sometimes 0.5 or more,
difficult to justify. 

The total injected energy of C1em3 is similar to the previous successful model,
a sign of a good self-regulation in the global feedback process.
This energy can be compared with the total energy radiated away, $\approx 1.38
\times 10^{60}$ erg (in the simulated half-space $z>0$). 
Interestingly, the present heating
provides more energy than that lost by radiation ($E_{\rm rad}\sim 1.4
\times 10^{60}$ erg).
Notice that, in every described model, the total outflow energy
(for the full-space system)
is about an order of magnitude lower than the 
available BH energy ($\sim1.8 \times 10^{62}$ erg).
The evolution of the energetics is very similar to the description
presented by G11. The core kinetic energy
increases after every intense AGN outburst, but it is soon dissipated
and transformed into potential energy through the expansion of the IGM.

\subsection[]{Intermittent feedback}

\subsubsection[]{Model Int510m: $10^{-4}P_{\rm Edd}$}

\begin{figure*}
\includegraphics[width=0.698\textwidth]{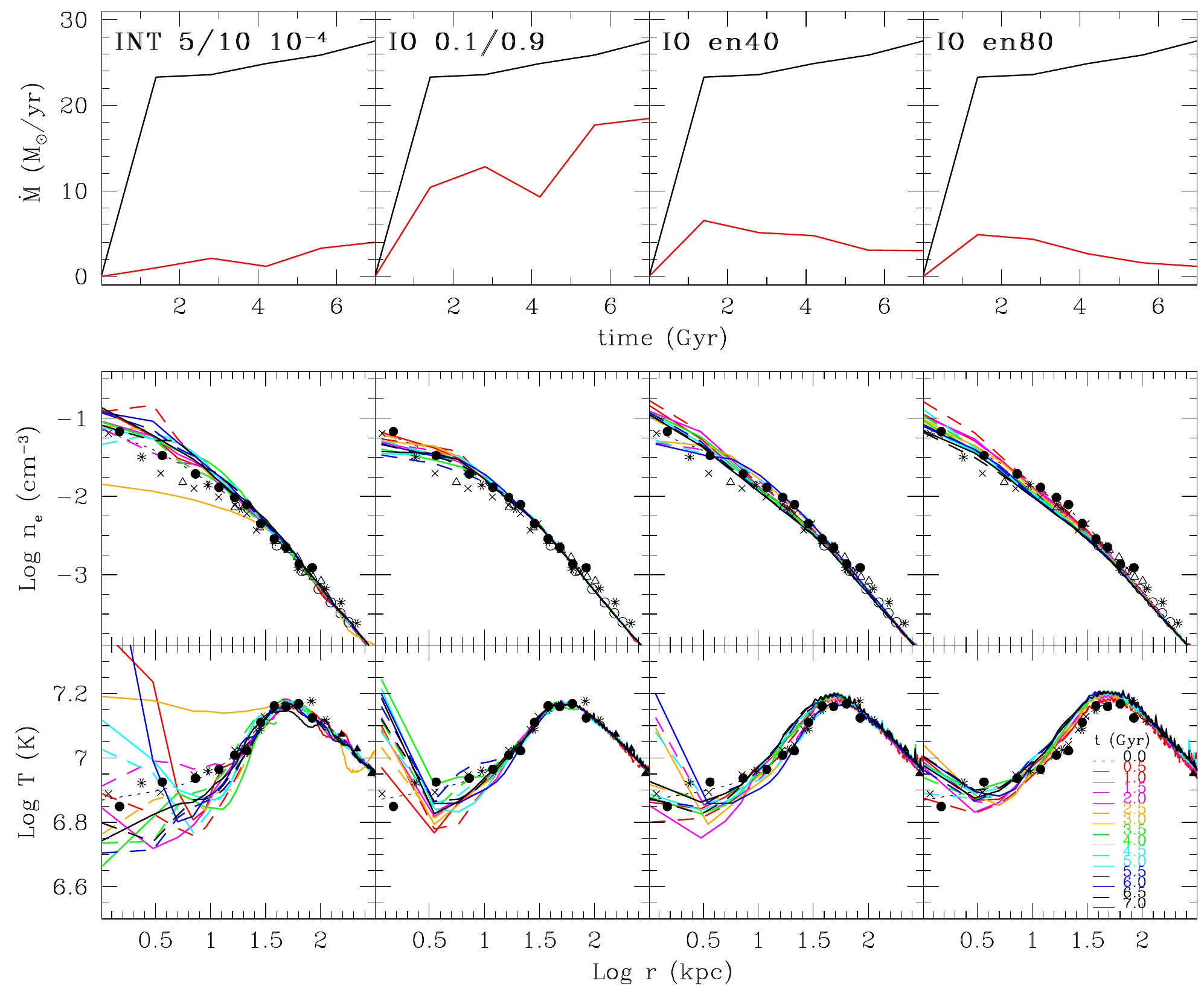}
\includegraphics[width=0.699\textwidth]{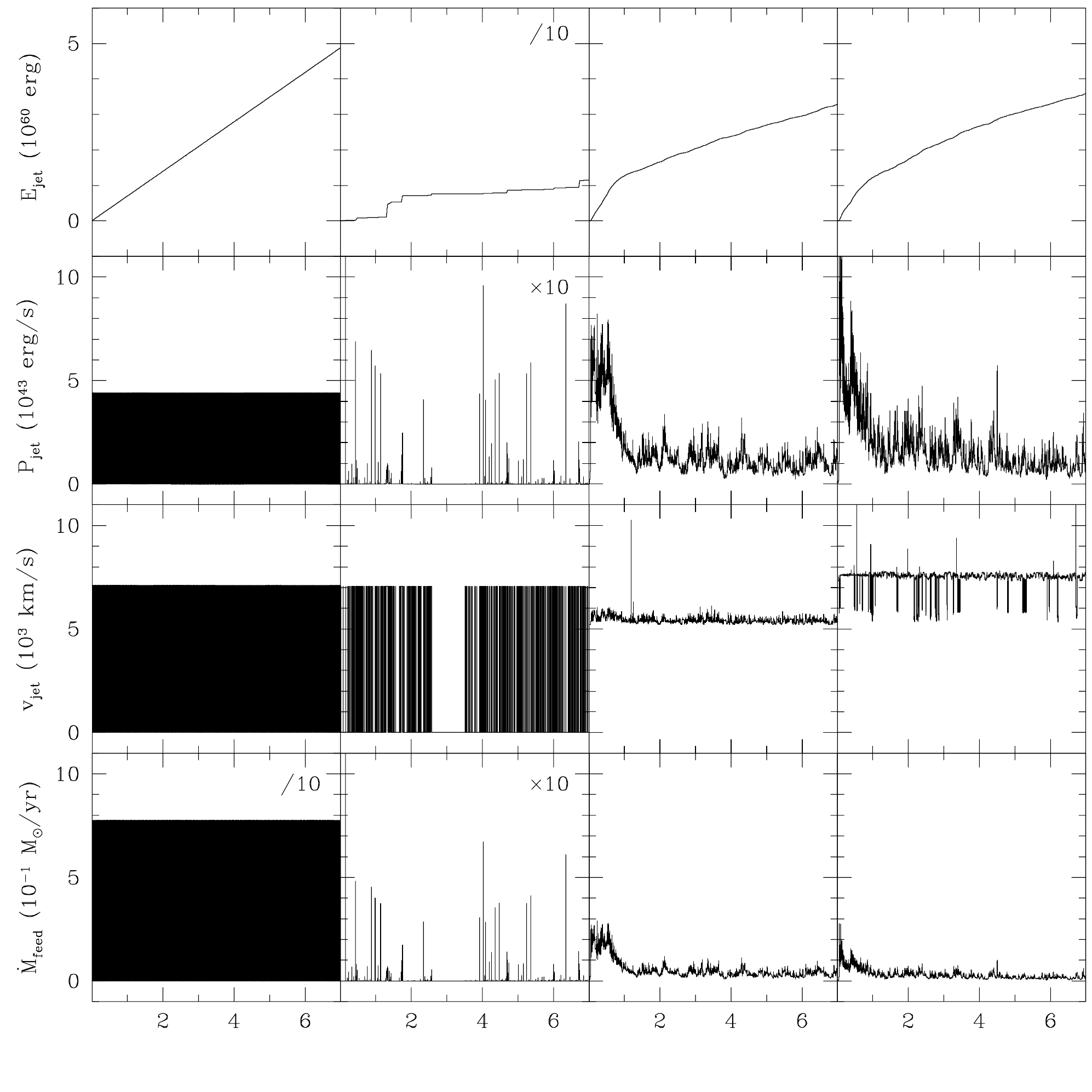}
\caption{Evolution of intermittent and InOut models (with or without mass loading). 
The description of the plots are same as for Fig. 2.} \label{fig:extra}
\end{figure*}

In this series of models we want to decouple the simulation from the
self-regulation engine, in order to control some important
AGN outbursts parameters. For example, it is interesting to fix
the frequency of the outflows.
In G11 we found that intermittent outflows, with jet duration
of 5 Myr and activated every 10 Myr, are very effective in inhibiting
gas cooling preserving the correct temperature and density profiles
in galaxy clusters. 
This AGN activity cycle agrees with observations of acoustic `ripples' in Perseus
cluster (e.g., \citealt{fab06}).

In the models described in this Section 
we do not insert an ad hoc velocity, but we retrieve
its value from setting a constant power, here taken equal
to $10^{-4} P_{\rm Edd}$.
Thus $P_{\rm jet} \sim 
4\times10^{43}$ erg s$^{-1}$. We choose the `nozzle' injection
method, therefore the velocity is constant: 7000 km s$^{-1}$.
Note that the efficiency is irrelevant in the computation.

As a result (Fig. \ref{fig:extra}, first column), 
the cooling rate is less than a few $\msun$ yr$^{-1}$,
for several Gyr, and shows a slow increase after 4.5 Gyr,
up to 4 $\msun$ yr$^{-1}$. The azimuthally averaged density
profiles are almost superposed to the observations, with a
slight decrease in time, becoming flatter.

Thus, like in Bondi models, the radiative cooling is
significantly reduced by the simulated heating process.
Using the nozzle injection (essentially an outflow with negligible
height) produce a spike in the very few kpc, while
at radii greater than 10 kpc the fluctuations from initial $T(r)$ 
are almost zero. As expected, shock heating is more vigorous in the nucleus
when the outflow is injected through the boundary, 
while more effective at intermediate radii,
when the entrainment method is adopted.

In this type of intermittent models it is not granted that the feedback
produces evident cavities. In fact with a cycle of 5-10 Myr the head 
of the jet cocoon almost touches the backflow tail of the last bubble.
Thus, we expect that with a duty cycle greater than 50 per cent
the jet becomes practically continuous.

\subsubsection[]{Model Int510h/l: $10^{-3}$ and $10^{-5} P_{\rm Edd}$} 

Keeping the same cycle scheme (5-10 Myr), but increasing the
power by a factor of 10 produces a typical `exploding' simulation (not shown).
The temperature profile gets a very steep negative gradient,
as soon as the outflow ignites, destroying the entire cool core
and generating a `plateau' in $n(r)$.
The cooling rates are obviously zero, but with unacceptable observables.

On the other hand, with a much lower power ($4\times10^{42}$ erg s$^{-1}$)
$\dot{M}_{\rm cool}$ approaches 12 $\msun$ yr$^{-1}$ after just 1 Gyr.
Generated bubbles are quite stable and, being slowly inflated, show
cold rims in the early phases. Nevertheless, the density and temperature
profiles resemble those of a classical cooling flow, with less peaked gradients.

We can conclude that, even with fixed intermittency, an acceptable AGN feedback model
should not approach an Eddington regime (like proposed
in some analytical works, e.g \citealt{kin09}), at least in galaxy groups, where
a strong burst can easily wreck the delicate thermodynamic structure of
the core.

\subsubsection[]{Varying fixed cycle: 1-10, 2.5-10, 3.3-10, 7.5-10}

Keeping a fixed jet power, as in our acceptable model Int510m, but
this time varying the duration of the AGN activity, leads to 
models that must be rejected.

Indeed, when every outflow event has a duration of 1, 2.5 or 3.3 Myr 
the intermittent
feedback becomes quite ineffective in halting the inflowing
cold gas: the central gas density
slowly grows with time until radiative cooling prevails over heating, causing the
cooling rate to surpass the threshold of acceptability.
For example after just 1 Gyr, with 3.3 Myr duration, the cooling rate
is over 7 $\msun$ yr$^{-1}$ and the temperature curve turns very similar 
to those of model CF.

On the contrary, when the jet duration exceeds 5 Myr
the inflated
cavities disappear, because the different outbursts melt rapidly
in a ten kpc zone, producing an almost continuous injection. The
consequence is a steep $T$ gradient at the center, which can halt
the cooling flow, but without self-regulation it easily overheats the
system.

\subsection[]{Thermal feedback}

\subsubsection[]{Adding thermal energy: $20\%$ up to $99\%$}

It is worth noting that one difficulty of a pure kinetic
outflow (often called `momentum-driven') is that it can easily carve
a tunnel in the surrounding IGM if the ram
pressure of the jet ($\rho_{\rm j}v_{\rm j}^2$) is much larger than
the thermal pressure ($\rho c_{\rm s}^2/\gamma$) of the gas.
Thus a feedback with high kinetic energy will advance undisturbed in
the group core, depositing its energy at intermediate radii ($15-30$ kpc).
At the base, the cold gas accumulates in a torus-like manner.
This is especially evident at early stages of a momentum-driven outflow.
The successful models, we found, are able to quench cooling flow, because
at some point the generated turbulence induces strong mixing in the gas,
also in the very center. Therefore the circulation inside the jet
path makes shock heating very effective. In our models
the turbulence is 
sustained only by the AGN feedback. It is likely that this process
dominates in the very central region of the system, 
although turbulent and bulk motions caused by
cosmological accretion, usually very subsonic, may also contribute
(e.g., \citealt{hei06}).

Nevertheless, the high density gas in the torus cools efficiently
and triggers the AGN feedback in an almost continuous way.
It is physically reasonable
that a part of the jet power is converted in thermal energy,
especially during the entrainment process. Based on these considerations,
we simulated some models
with a fraction of the total injected energy
in the form of thermal energy, through the nozzle.
We kept the Bondi prescription for the feedback method.

As pedagogical purpose, we first set the fraction of $E_{\rm th}$
as 99 per cent the total energy. 
This setup is similar to the model of
\citet{cat07} with loading factor of 100 and Bondi efficiency 0.1.
As expected, the first powerful outflow (over $10^{44}$ erg s$^{-1}$) 
destroys the entire thermodynamic structure
of the cool core: the produced shock (with temperatures
above $10^{10}$ K) is almost spherical.
This is another indication that the group
necessitates a much more delicate heating compared to a cluster.
Lowering the efficiency to $10^{-3}$ still generates a similar catastrophic
heating.

On the other hand, keeping the above efficiency, but reducing the
fraction of $E_{\rm th}$ to $20\%$ or $33\%$ results in models
overall similar to purely kinetic ones. One major difference are small
puffs at the base of the jet, that enable more turbulence. The problem
is that, in the inner $8-10$ kpc, temperature profiles show
a spike, due to injected $E_{\rm th}$. Another interesting effect is
the enlargement of the tunnel carved by the outflow, because
of the increased internal pressure, compared to the momentum-driven
one. 

The best model seems to be Eth50 (not shown), with $50\%$ thermal energy
and efficiency $5\times10^{-3}$. The feeble jet is almost
always continuous, with velocity 3000 km s$^{-1}$, but the thermal
energy injection
generates a break-up in the jet structure, which fragments
in small buoyant bubbles of size $5-8$ kpc; 
a very interesting feature that is seen in NGC 5044
(\citealt{dav09,dav10}). 

We conclude that a good fraction of thermal energy associated with
the AGN feedback can help the deposition of heating at
very small radii, reducing the cold torus at the base of the jet
and fragmenting the outflow in small size bubbles. 
At the same time, however, it is difficult to obtain a flat 
or positive (mass-weighted) $T$ profile. We will expand this analysis
in a future work, with very high resolution.

\subsection[]{In \& out feedback} 

\subsubsection[]{Model IO1-9}

One problem of self-regulated feedback models is how much
of the gas contributes to the growth of the black hole mass.
Needless to say, with a resolution of $\sim 500$ pc
we cannot consistently estimate the accretion rate onto BH. 
As noted in some works (e.g., \citealt{cat07,ost10}; G11)
part of inflowing gas onto a black hole could be thrown back by the
generated outflow, perhaps through entrainment.
Therefore only a fraction of the accreting mass, $\Delta{M}_{\rm acc}$,
might actually be captured by the black hole. From another perspective, we can 
see this as a reduced total efficiency. 

That being stated, we computed some models in which we attempt
to track the part ($f_{\rm in}$) of infalling gas that really increases
the BH mass. The residual fraction ($f_{\rm out}=1-f_{\rm in}$) 
is considered the active mass of the outflow:
\begin{equation}\label{inout}
\frac{1}{2}f_{\rm out}\dot{M}_{\rm acc}v_{\rm jet}^2=f_{\rm in}\, \epsilon \dot{M}_{\rm acc} c^2.
\end{equation}
Note that in these models it is better to use the nozzle injection mode, in order to totally
control the outflowing mass, without altering the internal domain.

An interesting feature of this method is that the outflow velocity is
practically constant, dependent on $f_{\rm in}/f_{\rm out}$ (and $\epsilon$).
An observational constrain of the outflow velocity may thus give hints about this
ratio between infalling and outflowing matter. 

Setting $f_{\rm in}=0.1$ ($f_{\rm out}=0.9$) and efficiency $10^{-3}$,
the outflow, coming out of the nozzle, has a velocity around 7000 km s$^{-1}$.
With cold accretion, the power of the jet presents strong oscillations
between $10^{43}$ and $10^{45}$ erg s$^{-1}$ while
the cooling rate approaches in a few Gyr the pure CF simulation.
Even with Bondi accretion (same efficiency; Fig. \ref{fig:extra}, second column), 
which, in theory, is more regular, 
the (high) outflow power can not find 
a stable balance in time, resulting in $\dot M_{\rm cool} \sim 18$ 
$\msun$ yr$^{-1}$ at 7 Gyr. 
While some accretion events reach 4-5 $\msun$ yr$^{-1}$,
the produced strong feedback does not last long enough. In fact
the total injected energy is very low: $10^{59}$ erg.
It is interesting that the radial profiles of this model (IO1-9)
show a cool core, with the exception
of a central negative temperature gradient.
Therefore, overall acceptable mean $T$ and $n$ profiles do not necessary imply that
the system is not cooling.

Trying to increase the efficiency is a serious problem for the InOut model, because
the velocity will constantly overtake 50000 km s$^{-1}$. We could 
lower $f_{\rm in}$ in order to increase the jet mass, but with just
1 per cent of accreting mass (probably unrealistic) the velocity
will still have very large values.

\subsubsection[]{Model IOen40 and IOen80}

The only solution, we have found, is to return back to the `entrainment hypothesis'
and multiply $f_{\rm out}$ by a factor $\eta$ (like a mass loading
factor).
Following this assumption, with $\eta = 40$ and 
$f_{\rm out}=0.3$,
we could enhance the Bondi efficiency  ($\epsilon= 5\times10^{-3}$) to 
obtain an acceptable cooling rate, below 5 $\msun$ yr$^{-1}$ (Fig. \ref{fig:extra},
third column),
after the small bump at 1.5 Gyr. Now the velocity has a value of
$\sim6000$ km s$^{-1}$. With $\eta=80$ the average cooling rate is further reduced,
asymptotically decreasing to 1 $\msun$ yr$^{-1}$ (fourth column). 
Both simulations generate some of the best radial profiles, which
keep the cool core appearance for the entire evolution time, almost
`glued' to the initial (observed) condition.
This is an indication that the feedback has returned stable
(around $P_{\rm j}\sim 1-3\times10^{43}$ erg s$^{-1}$),
gentle and continuous, very similar to successful Bondi models (Sec. 3.2). 
Notice that the accretion rate is again sub-Eddington.

Overall, a pure InOut model exhibits a spasmodic behaviour, because of the intrinsic
linking between accreting and outflowing matter, which
can be smoothed out only by introducing a mass loading factor. However, this parameter
is nowadays unknown and therefore its value is chosen ad hoc, in order
to retrieve reasonable jet velocities. It is therefore simpler and cleaner to adopt
the usual entrainment feedback of the main models (Sec. 3.2 - 3.3), seeing that
the results are analogous.

\section{Dynamics and Observables}
\begin{figure*}
\includegraphics[width=\textwidth]{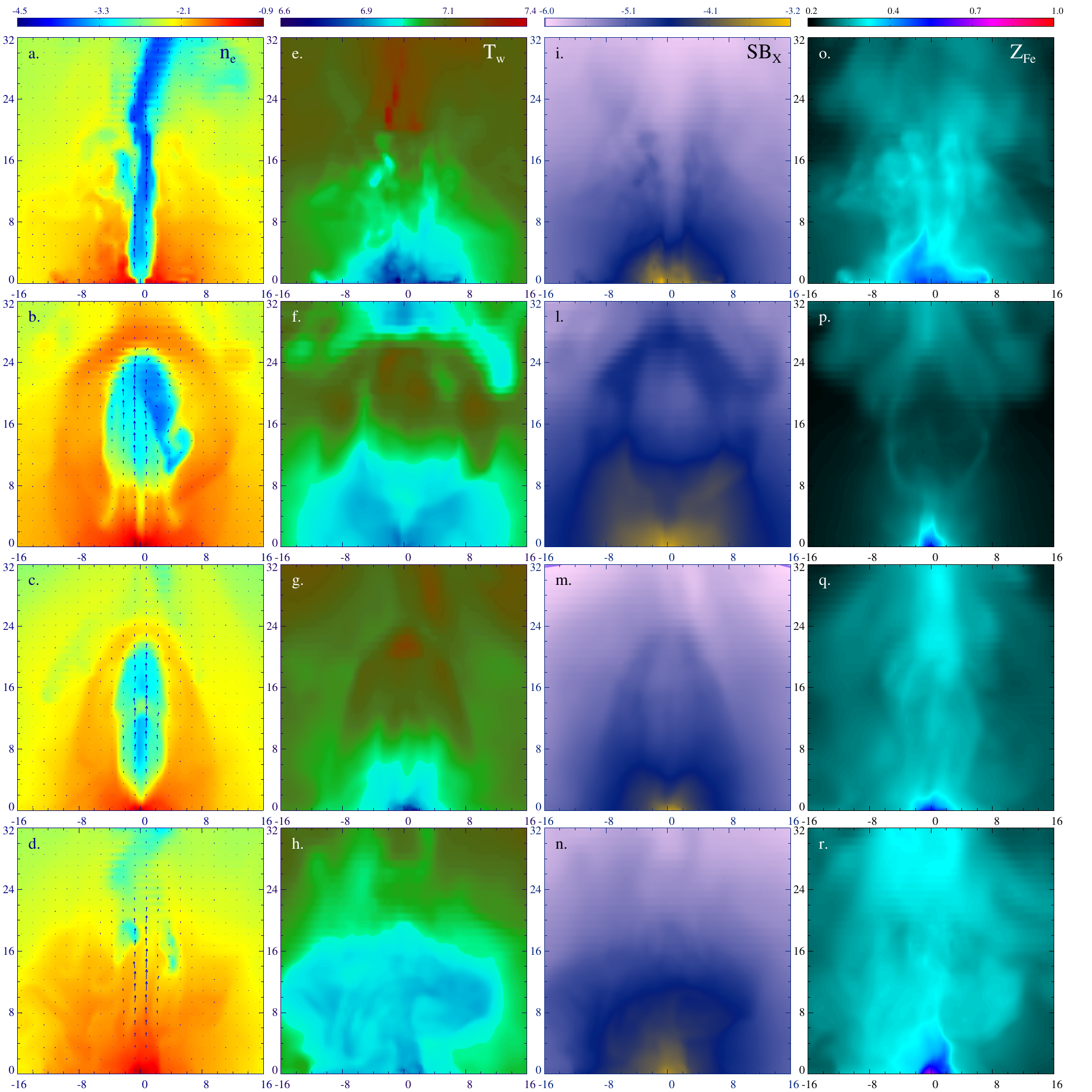}
\caption{Maps of four best models (kpc unit).
Columns, from left to right:
cuts through $x$-$z$ mid-plane
of electron number density (cm$^{-3}$, logarithmic)
with velocity field superimposed; emission-weighted temperature
maps (K, logarithmic); X-ray surface brightness maps
(erg s$^{-1}$ cm$^{-2}$); emission-weighted iron abundance maps
(Z$_{\rm \odot}$ unit). Rows, from top to bottom, are
associated with different models:
Bc5em2 (5 Gyr), C1me3 (0.5 Gyr), Bi5em2 (7 Gyr), Bi1em1 (6.5 Gyr).
The colour scale is given by each bar at the top.
} \label{fig:maps}
\end{figure*}
In this Section we will show the `real face' of the best
simulated models, at some relevant times. 
We will analyse physical quantities,
such as gas density, and other astronomical observables,
i.e. integrated emission-weighted maps. Notice that the
entire structure of our computation favours the study
of a very long term evolution. In order to fully comprehend 
the dynamics of single burst events, we should carry on
a short term evolution with much higher resolution. We will
confront this setup in a future dedicated paper.

\begin{figure}
\centering
\includegraphics[width=57mm]{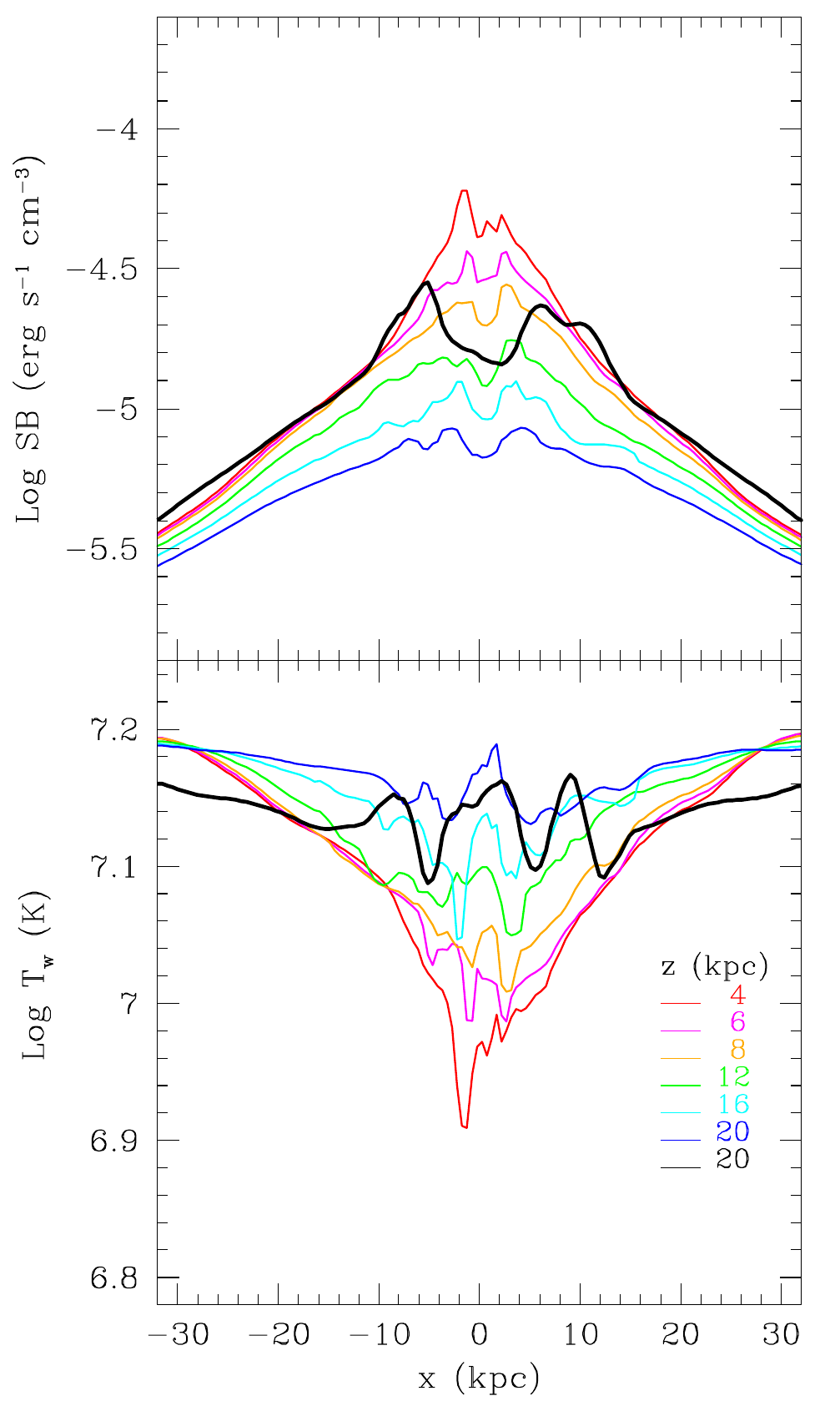}
\caption{X-ray surface brightness and emission-weighted temperature
1D cuts through $z$ (see legend). Black thick lines correspond to the model Bi1em1 
(panels n and h of Fig. \ref{fig:maps}), while the others belong to model Bc5em2 (panels i and e).
} \label{fig:1Dcut} 
\end{figure}

In Fig. \ref{fig:maps} the first column exhibits two-dimensional
density cuts trough the $x$-$z$ mid-plane (panels a-d)
with the velocity field superposed. The first
row is associated with Bc5em2 model (at 5 Gyr), one of 
the successful models in our set of simulations. 
This snapshot depicts one of the three typical
stages of the AGN-regulated simulation in a galaxy group. The gentle
and continuous heating of the outflow ($\sim10^{44}$ erg s$^{-1}$)
carves a narrow tunnel
in the intragroup medium with electron density around 
$2\times10^{-4}$ cm$^{-3}$. 
The diameter of the channel is usually 4-5 kpc,
collimated in the inner region by the thermal pressure
of the surrounding gas. At 14 kpc, where the jet ram pressure
becomes comparable to the thermal pressure 
of the ambient gas 
($P_{\rm th}\sim2.5\times10^{-11}$ erg cm$^{-3}$), the outflow structure
becomes perturbed and
more turbulent\footnote{
At 24 kpc the outflow is almost stopped by the IGM, generating a rise
in temperature, $\sim1.55\times10^{7}$ K (see also Fig. \ref{fig:1Dcut}).}. 
In fact, near the tunnel,
instabilities are visible, a signature of further IGM entrainment
and turbulent mixing. The latter has a relevant role in 
the feedback process, especially at later times, when the AGN
has been active for several Gyr (Fig. \ref{fig:Bondi_cont}, cooling
rate decreases with time). In fact, even if the turbulent mixing does not directly heat
the gas, it greatly favours the deposition of 
energy at the base of the jet, where the cooling
tends to dominate.

The fact that the global
cool core appearance is not greatly modified, is another
key result. Although the continuous outflow generates a tunnel,
the overall initial structure is conserved. 
On the other hand, increasing the
efficiency of Bondi models leads to a wider V-shape channel, which
begins to dominate the core of the group. This is not observed
and in fact those models are rejected also for the central
negative $T$ gradient. As seen in the second column 
(panel e), the (emission-weighted) temperature of 
Bc5em2 has instead a positive
gradient, almost flat in the 5 kpc nucleus. 

The continuous presence of the channel (interrupted sometimes
by the fragmentation) is a feature not commonly seen
in real groups, at least in the local universe\footnote{The
evolution of groups is still far to be certain, lacking
a complete sample at intermediate and high redshift.}.
However, in the third column (panel i) the X-ray surface brightness map does not
show an evident channel, with only two very faint features (at $12-16$ kpc), 
similar to a pair of  `arms'.
Going into quantitative details we performed one dimensional (1D)
cuts through $z$ of SB$_{\rm X}$
at different levels (from 4 up to 20 kpc; see Fig. \ref{fig:1Dcut},
non-black lines). It is clear that the central depressions
are not deep ($15-20$\%, usually with a difference of $5\times10^{-6}$ erg s$^{-1}$ cm$^{-3}$), 
plus the width is particularly slim (few kpc), making
the detection of the channel extremely difficult\footnote{In addition, Poisson noise, background,
resolution, and response of the X-ray detector (not present in our mock maps)
could easily alter or obscure this very faint and
narrow feature.}. Over 20 kpc the
slim tunnel is lost in the integration trough line of sight.

The last column is associated with the tracer of the iron
abundance (again emission-weighted), injected
by SNIa and stellar winds of the cD galaxy (see G11 for details).
This type of advected quantity is deeply linked to the history
of the AGN heating. In panel o is indeed possible to recognize
the outflow pattern, which is asymmetrical and can drag the
metals produced deep inside the central elliptical galaxy 
up to 15-20 kpc, with values
around $0.3-0.4$ Z$_{\rm \odot}$. This characteristic is
commonly observed in the core of AGN heated systems 
(e.g. \citealt{kir09,kir11,rap09,dav10}; Doria et al., in preparation).

The second row of Fig. \ref{fig:maps}, 
shows another good model, one
based on the cold feedback.
C1em3 (at 0.5 Gyr) shows the other face of the AGN-regulated
evolution: an AGN burst with power
$\sim 1.2\times10^{45}$ erg s$^{-1}$ generates a big
cavity in the IGM density, with major/minor axis of 16/10
kpc. The injected energy is almost $10^{60}$ erg, with an
outflow velocity over $10^4$ km s$^{-1}$. 
The buoyant bubble has a high 
density contrast with the environment, $\sim100$.

In contrast to the violent cavities generated in our
previous galaxy cluster simulations (G11), the cold accretion
mechanism is able to inflate low density cavities without
heavy shocks. The outflow power is in the group simulations several orders of magnitude lower
and the injection per timestep is usually smoother. 
In fact (see panel f and Fig. \ref{fig:1Dcut}, black line of bottom panel),
$T_{\rm w} \sim 1.45 \times 10^7$ K inside the shell
which is
only slightly larger than the surrounding gas value.
This is also suggested by the homogeneity of the map above 12 kpc
(reddish colours).

A striking feature, usually hard to detect, is the bright rim, that
surrounds the cavity (panel l). The rim, especially the low-$z$
region, is formed by low-entropy gas originally in the center.
This feature is often seen in deep
X-ray observations (e.g. \citealt{sal06} and references therein), 
suggesting that the rims are colder than
the average ambient medium. The 1D cut through $z=20$ kpc 
(Fig. \ref{fig:1Dcut}) clearly confirms the presence
of this kind of rim, a drop of temperature (about 40\%) coincident with the high
X-ray emission (at $x=\pm6$ kpc). 

In panel p we show the (emission-weighted) iron abundance map
in the $x-z$ plane. The iron-rich core, few kpc in size, is clearly
visible. At $z\sim 28$ kpc there is a region of Fe-rich gas, lifted
by the outburst. The dense cavity rims also have a slightly larger
abundance than the ambient gas, also revealing that the origin of a part
of the rim material is connected to the nucleus of the group. 

The SB$_{\rm X}$ inside the cavity (panel l) exhibits 
a depression of $\sim50$ per cent with respect to the rims.
Notice how this feature (Fig. \ref{fig:1Dcut}, black line) would be
easily detectable in the X-ray, compared to the vanishing faint channel,
due to its deep depression and large width.  Moreover, the
nucleus (5-8 kpc) dominates the emission, while the upper
part of the bubble vanishes rapidly in the background ($>30$ kpc).
The future of this bubble is to buoy outwards and being 
destroyed, after few tens Myr, by the backflow and instabilities.

The third row presents the snapshots of model Bi5em2 at 7 Gyr.
At that stage of evolution the IGM has accumulated more 
turbulence, after several AGN events, promoting the diffusion
of iron in a radius of 20 kpc from the center (panel q).
However, this particular moment is again dominated 
at the bottom-center by another AGN outburst ($2\times10^{43}$ erg s$^{-1}$).
Therefore, the young bubble (panel c) still buoys undisturbed in the IGM, generating
a slight asymmetry in the iron distribution along jet axis, over the more uniform background. 

At this time we caught indeed the early
injection phase of the outflow. A quasi cylindrical cocoon 
envelops the core of the jet, which has now a velocity around
3000 km s$^{-1}$. The injected initial velocity was over $10^4$ km s$^{-1}$,
meaning that the jet commonly decelerates in a rapid way after few kpc. 
At $z=22$ kpc the emission-weighted temperature reaches the maximum
value of $1.62\times10^{7}$ K (panel g), clearly indicating the
effect of shock heating.
Around the contact discontinuity the gas is weakly shocked, with
temperatures above $1.25\times10^{7}$ K. The average Mach
number is around 1.2, a typical value found by observations
(\citealt{bla09,git10}).
The arc is also visible in
the SB$_{\rm X}$ map (panel m), while the decrement 
associated with the cavity is
$\sim35-40$ per cent.

Notice that the maps of this model are very similar to
those of C5em4, another successful model, 
which has analogous bubbles production. This behaviour
confirms further the strict relationship between
(intermittent) Bondi and cold models, as previously noted
by the compared efficiency analysis.

The last row of maps displays another snapshot, typical for
intermittent (self-regulated) feedback models, like Bi1em1.
The big outburst and the generated bubble have been vanished,
leaving the system in the aforementioned turbulent phase,
in which mixing distributes the low residual heating to the central
IGM. Soon after this period, the gas will cool and start the inflow.
The velocity
field is relatively 
chaotic (panel d), with maximum velocity of 640 km s$^{-1}$,
a bit lower than local speed of sound, evidently decreasing 
in the turbulent zones. 
The mixing of gas at this epoch restores the spherical symmetry of the cool core, 
with average temperature of 
$\sim9\times10^{6}$ K (panel h). This is underlined also
by the symmetric surface brightness map (panel n) and
by the substantial diffusion of Fe in the intermediate zone, $10-30$ kpc (panel r).
Note that the abundance radial profile at the very center is, 
however, more peaked
during this quiescent cooling flow phase ($\sim 0.8$ Z$_{\odot}$),
compared to the AGN outburst period. Thus, our AGN models
might also explain the dichotomy of cool core and non cool core
groups, having the former higher central abundances
(\citealt{joh11}). 

An interesting feature, captured at this stage
(often present after an AGN outburst), is the sequence
of very small `bubbles' at $r>15$
kpc, with typical diameter of a few kpc. 
The origin of this tiny `bubbles' is 
the fragmentation of the jet due to turbulent and chaotic
motions, previously discussed. In this map they
are in a very late phase, almost disappearing in the ambient
medium. In this case they would be probably not detected in the X-ray
(just a 10 per cent jump in surface brightness). To capture their signature,
an observation should be done at an earlier evolutionary moment,
after a big event, when the jet is rapidly collapsing.
Nevertheless, it is striking that \citet{dav09,dav10} 
find exactly this type of configuration in NGC 5044:
few big cavities and many small `weather-driven' bubbles,
which appear to be radio quiet (i.e. not directly inflated
by the jet).\\

Overall, we conclude that the AGN outflows dynamics is
particularly complex, as expected from the previous study
in galaxy clusters. The main stages of evolution for
the best models are: 
(i) a gentle continuous outflow, which sustains the cooling
flow for the majority of time (or always in Bc models)
and which is barely observable in SB$_{\rm X}$ maps;
(ii) a phase of large cavity inflation, with cold rims
and high density/SB$_{\rm X}$ contrast (only for
non-continuous models);
(iii) small periods of quiescence in which the turbulence
and vorticity promotes mixing and tiny
weather-driven `bubbles'. 

The presence of the above described features
(big and small cavities, asymmetry of Fe abundance, 
low Mach shocks, etc. ), is
consistent with several recent X-ray observations of
heated systems, and demonstrates
the fundamental role of AGN driven outflows in the 
evolution of groups (and clusters).

\section[]{Discussion}
In this paper we have proposed 3D
numerical simulations of self-regulated
AGN outflows in a galaxy group environment.
We have followed the long term evolution of the
interaction between two main characters: 
radiative cooling and feedback heating. The aim of this work
is to understand the general features of the feedback 
process to solve the cooling flow
problem, i.e. quenching cooling rates, but, at the same time, 
preserving the global cool core structure.

We tested two relevant models, which we had found to be successful
in galaxy clusters (\citealt{gas11}): cold feedback and Bondi feedback
mechanisms. Galaxy groups, with masses $\sim$ two order of magnitude
less than big clusters are more fragile systems and the feedback
must be consequently more delicate.
In the following we discuss and summarize the main results
obtained, along with merits and flaws of the models.

\subsection{Hot continuous Bondi (Bc)}
In the first set of simulations the commonly adopted Bondi
prescription is applied to evaluate the accretion rate
onto the black hole.
Although the Bondi
scenario is quite naive, it can be treated as the template model for
a typical self-regulated feedback directly linked to entropy.
The efficiency, as free parameter, might just incorporate
the discrepancies between the ideal Bondi model and the
real dynamics.

Overall, we found (similarly to clusters) that 
the Bondi accretion models are able to satisfy
our strict requirements for a plausible evolution, with
a mechanical efficiency
in the range $5\times10^{-2}-10^{-1}$. 
As expected,
the accretion rates are always two to three orders of magnitude
lower than Eddington limit. This implies that the outflow `touch'
is always delicate and non-explosive, with power of the order 
of $10^{44}$ erg s$^{-1}$ and velocity $10^4$ km s$^{-1}$.
These values are consistent with numerous observations
of AGN outflows through absorption lines (e.g. \citealt{nes11}).

In the best models cooling rates are reduced below a 
few or fraction of $\msun$ 
yr$^{-1}$, with a monotonic decrease in time. The density radial
profiles of these models do not deviate too much from the observed ones. The same
can be said of the temperature gradient, with a small
overheating in the inner 4 kpc nucleus. 
The last feature is probably associated with the continuous nature
of the feedback: the Bondi formula is in fact
based on the accretion of hot gas, whose
reservoir never drains.

A problem for this model may be the continuous injection of energy,
which carves a tunnel in the IGM. However, as shown in Section 4,
this channel is usually very narrow and almost disappears in 
the surface brightness image (on the contrary of events associated with big bubbles). 
The important result is that
a feeble continuous heating could operate, without
being in contradiction with observations.

The creation of cavities of tens kpc is another issue. 
However, we have noted that, during the increasing
generation of turbulence (due to the AGN), the jet becomes gradually
more disturbed: eventually it will fragment, producing a series
of small buoyant bubbles. This phenomenon is certainly enhanced
by a cosmological evolution through merging and other large scale
motions (e.g., \citealt{mor10}).

We have also checked that slightly different initial conditions, like
an initial isothermal temperature profile and a NFW dark matter
profile, do not alter the behaviour of the feedback, producing
similar results.

\subsection{Cold feedback (C)}
In galaxy clusters the triggering mechanism linked to the
instantaneous $\Delta M_{\rm cold}$ induced a very powerful
outburst, often super-Eddington. In much less massive
galaxy groups
such a powerful feedback would easily destroy the
thermal structure observed,
producing negative temperature gradients
up to 100 kpc.
Therefore, successful models adopt efficiencies about
an order of magnitude lower with respect to those used
for clusters.
With $\epsilon_{\rm c}$ between $5\times10^{-4}$ and $10^{-3}$ cooling rates
are still well below 10 per cent of the pure CF model, but
now with consistent central densities and temperatures.

The model is naturally impulsive, 
but this time the duty cycle is very high compared
to the cluster evolution shown in G11. In fact, for the majority
of time the outflow is continuous with moderate power, 
similar to the previous Bondi models. 
The AGN stays in a quiescent phase only 15-20 per cent of the 
time, after which it will ignites 
with outbursts up to $10^{45}$
erg s$^{-1}$. These powerful jets are essential for creating
big bubbles of 10-20 kpc diameter.

One of such event has been analysed 
in Section 4 (Fig. \ref{fig:maps}). The bubble does not
show high temperature within (e.g.
$10^8$ K or more), like in clusters (G11):
it is very under-dense with respect to the ambient medium,
with a relatively large contrast in the SB$_{\rm X}$ map ($>50\%$). It is striking
that the metal abundance is often highly asymmetric, enhanced
along the outflow direction.
All the previous features are commonly present in deep 
X-ray observations (\citealt{mcn07,bal09,gast09,kir09,kir11,rap09,dav09,dav10}).

Overall, it is surprising that, in galaxy groups, models with cold feedback
resemble, most of the time, simulations with Bondi accretion, with
the merit of adding important features like X-ray cavities and
ripples at larger radii (essentially weak shocks with Mach around
$1.1-1.3$). This fact clearly indicates the necessary properties
of the feedback process, if outflows are the main mechanism:
very frequent or almost continuous and gentle.

\subsection{Bondi with cold timing (Bi)}

Simulations using the
third class of models, Bi, have been carried out
to test the combined effect of Bondi accretion and cold triggering
mode, in order to avoid Bondi continuous injection (intrinsic in the ideal case). 
In some preliminary tests (here not shown), we found that a simple superposition
of the two feedback schemes leads to 
unacceptable models, because the outcome is just a
more powerful jet, which carves a larger and deeper continuous channel.

Therefore, we tried another way to combine both methods. 
Keeping Bondi accretion prototype (entropy-regulated) for the value of power,
we assumed that when the gas begins to cool at the center
(i.e. the inflow initiates), the feedback
is activated. 
This will prevent too peaked instantaneous outflow
powers and, at the same time, the AGN activity acquires a sort
of duty cycle.

The results are very encouraging. The required Bondi efficiency
for best models stays again in the range $5\times10^{-2}-10^{-1}$.
Assuming $\epsilon_{\rm B}<10^{-2}$, the evolution
manifests the typical flaws of a pure CF run. On the contrary, if
$\epsilon_{\rm B}>10^{-1}$ the feedback becomes very explosive and easily `burns'
the cool core. The best Bi models show instead radial profiles
with gradients not far detached from the initial observed state,
along with very low cooling rates (usually one or less
solar masses per year).

The look of the Bi simulations has been analysed, in-depth,
in Section 4. The general features are almost identical to the
previous discussed models. Extensive periods of a continuous
Bondi-like injection ($P_{\rm j}\sim10^{44}$ erg s$^{-1}$,
$v_{\rm j}\sim10^4$ km s$^{-1}$) 
produce a narrow heating channel, alternated
with dormant AGN phases ($\sim15\%$). As in the cold feedback
computations, the short inactivity promotes the subsequent
generation of outbursts with greater power (3-4 times more than
the continuous phase), producing important observed features,
such as tens of kpc cavities and oscillations in the flow
quantities.

It is relevant to note that, as in previous good cold models,
only a fraction
of the injected mechanical energy is used
to inflate cavities (usually under 50 per cent\footnote{
E.g., the inflation energy of the bubble in Fig. \ref{fig:maps}, panel b,
is $4PV\sim4\,(2\times10^{-11}$ erg cm$^{-3})\,(1.1\times10^{68}$
cm$^3)\approx9\times10^{57}$ erg,
while injected mechanical energy is $2.45\times10^{58}$ erg. Thus, inflation energy
is just 37\% of the total injected energy in that single event.}).
One of the main results, also pointed
out in G11 for clusters, is that the success of a feedback
crucially depends on how the energy is transferred to the
surrounding medium, not only on the total amount of available
energy. Furthermore, the heated flow
has a very complex dynamics, with numerous facets.
After the mechanical energy is injected in the system,
a part of energy inflates the bubble, but the rest generates
the cocoon and weak shocks. When the jet is more continuous, the
moderate energy released at intermediate radii induces
a turbulence growing in time. These chaotic motions permits a more
efficient circulation of the hot gas, promoting the deposition of energy
through mixing at the
base of the jet, where the cooling tends to dominate.
Furthermore, while the
system becomes more turbulent, the feeble jet can be fragmented
and can cyclically produce tiny `bubbles', few kpc in size (albeit very faint in the SB$_{\rm X}$ map).
Their buoyant motion may promote additional mixing and heating,
helping the global feedback process. 
Quite simple feedback models still generate
an amazing variety of heating processes.

As a concluding remark for these models,
the efficiency comparison underlines again the strict
relationship with the cold best models, exactly around
$\epsilon_{\rm c}\sim5\times10^{-4}-10^{-3}$. 
We can therefore affirm that the feedback in galaxy groups,
able to quench cooling flow but at the same time preserve
the cool core structure, is globally only of one type:
moderate, gentle, often continuous and turbulent, 
with very short periods
of inactivity and following outbursts.    

\subsection{Rejected models}
It is also interesting to highlight the main
features of unsuccessful models, in order to understand
why some kinds of feedback do not work.

First of all, while a small part of the total energy injected
could be in thermal form, above a threshold fraction 
of $\approx 50$ \% the center of the group  becomes too hot with
respect to the observations (see also Brighenti \& Mathews 2002, 2003).
Injection of a modest amount of thermal energy 
can have the beneficial effect of disrupting the otherwise
ordered structure of the outflows (Section 3.5). This model intermittently
eliminates the long tunnel, usually not observed in real systems.
Moreover, the deposition of
heat in the nucleus helps preventing the formation of a cold torus.
Injection of thermal energy can be justified
assuming that shocks heat the hot IGM in the very central
region.

Another rejected model considers the active outflowing
mass directly linked to the real accreted mass onto BH
(InOut feedback, Section 3.6).
The main problem here is that, even with very low $f_{\rm in}$
(e.g. $f_{\rm out}>0.9$), the jet velocity often exceeds
$10^5$ km s$^{-1}$. This unappealing result can in principle
be fixed
adding a mass loading factor ($\eta f_{\rm out}$; see
also \citealt{cat07,ost10}). However, this assumption will mimic entrainment,
just as our standard outflow generation method described in Section 2.2.
It is not surprising therefore that we recover acceptable results,
similar to those described in Sections 5.1.

Intermittent models (Section 3.4), with fixed
jet power, are only partially acceptable, as was also found for
galaxy clusters (G11). The most successful run has relatively weak
outflows ($P_{\rm jet}\sim 10^{-4} P_{\rm Edd}$, activated
every 10 Myr, with a duration of 5 Myr each (model Int510m).
We have also tested jets with Eddington power
($\sim10^{47}$ erg s$^{-1}$). As expected they will
destroy the cool core generating systems very different from the real
ones.

Probably the major flaw of the intermittent models 
is their artificial
nature. Cooling rates can be suppressed, compared to a
pure CF model, but the temperature profiles oscillate
strongly in the center. 
This further supports the common notion that heating {\it must} be
self-regulated.

Finally, we have briefly investigated the effect of jet
precession (not shown). Important parameters like
inclination angles and revolution time of the jets are currently
poorly known. Nevertheless, we found
that the behaviour of cold gas at the base of the outflow,
i.e. the formation of a small cold torus, is not greatly
different from non-precessing models.
In fact, the heating channel will be again
generated after the first episodes, getting a tilted
orientation in respect to a normal $z$-axis cylindrical simulation.

\subsection{Comparison with galaxy clusters}

As expected, our simulations show that
AGN heating has a deeper impact
in galaxy groups than in clusters, 
because the same mechanical power per particle
has a greater effect for a lighter, less bounded systems.

As we have seen in the previous Sections, a jet
with Eddington power can easily erase the entire thermodynamic
structure of the cool core. The same applies for all the 
self-regulated simulations with very high efficiencies 
($>5\times10^{-3}$
for cold models; $>10^{-1}$ for Bondi). In G11 we showed
that consistent cold models commonly produced super-Eddington
outbursts ($>10^{47}$ erg s$^{-1}$), with a very
low frequency ($\sim10$ per cent).
Despite the huge power injected,
the shock-heating phase could not destroy the observed 
density and temperature gradients, and the cluster quickly
restored the usual cool core. 
The only marked consequences were more
ripples at larger radii and cavities, which in the
early phase presented high internal energy and shocked rims.

On the contrary, galaxy groups, with much less binding energy (per particle),
can not recover from the same strong impulsive feedback.
In order to prevent such
`heating catastrophe', cold feedback models in galaxy groups
require an efficiency at least $5-10$ times lower. 
The outcome is striking, because the evolution
resembles that of Bondi models, with low
power jets ($10^{44}-10^{45}$ erg s$^{-1}$) activated in a  
quasi-continuous gentle way. 
These outbursts occurring
after short quiescence periods (tens Myr) generate
$\sim 10$ kpc cavities surrounded by relatively cold rims
in the region close to the center of the group,
similar to some observations (e.g., \citealt{gast09}). 
Unlike in galaxy clusters, cavities in groups present
the usual low density contrast, with
temperatures similar to the 
surrounding medium. 

The best models assuming continuous Bondi accretion,
on the other hand, are equivalent in both
clusters and groups (except for the efficiency, lower in groups). 
Accretion rates always follow a sub-Eddington
regime. The moderate outflows carve a narrow tunnel in the
intergalactic medium; we showed, in this paper, that
it is very faint in the SB$_{\rm X}$ map and thus hard to resolve.
The absence of inflated bubbles
is still a riddle of this type of feedback.
Notice that in galaxy clusters the low-power outflows 
($\epsilon_{\rm B}=0.1$) had more difficulty
in halting the massive cooling flow (tens $\msun$ yr$^{-1}$),
showing more peaked density profiles. Therefore, in clusters,
cold accretion was slightly favoured as an acceptable model.

Another common feature in both clusters and groups is the
asymmetrical distribution of metals (mostly iron)
produced by the SNIa in the
central galaxy. They tend to accumulate along the jet axis,
in qualitative agreement with recent observations
(see Sec. 4).

In all the consistent group models the sub-Eddington outflow powers 
imply that the total injected energy (a few $10^{61}$ erg for the
full-space system) is always an order
below the total `available' BH energy. In fact,
during the simulated evolution, the black hole mass
increases at most 10 per cent
in Bondi models. In the cold feedback model the lower efficiencies imply a
much higher $\Delta M_{\rm BH}$, similarly to galaxy clusters. 
As noted in G11, however,
the real efficiency (still unknown) could just be higher, 
with a lower mass actually falling onto the BH.

Our simple approximation of the real sub-pc accretion does not allow
to answer why the efficiency in groups should be a factor 5-10 smaller
than in clusters, aside the observational requirement. 
It would be interesting to investigate this topic with
a dedicated analysis.



\section[]{Conclusions}

Here we summarize the main conclusions of our study.

i) Feedback triggered by both Bondi accretion and
cold mode accretion (or a combination) lead to successful self-regulated 
models. These schemes are able to quench cooling flow for many Gyr,
preserving the observed thermal structure of relaxed groups.
The required mechanical efficiencies
for cold accretion models are in the range $5\times10^{-4}-10^{-3}$,
while $5\times10^{-2}-10^{-1}$ for the Bondi scenario.\\

ii) In the galaxy group, the two main feedback schemes
generate similar flows. The global evolution is dominated
by almost continuous outflows, with sub-Eddington power 
($10^{44}-10^{45}$ erg s$^{-1}$) and velocities around
$10^{4}$ km s$^{-1}$. These values are consistent with those based
on observations of AGN outflows
(\citealt{cre03,mor05,mor07,nes08,nes11}).\\

iii) The main feature of the continuous phase is a 
narrow tunnel (large a few kpc, up to 30 kpc long), which
is almost undetectable in the SB$_{\rm X}$ map.
The tunnel is periodically
fragmented by the increasing AGN-driven turbulence, especially when
the BH is in a phase of low accretion.
The products are tiny, kpc-size buoyant bubbles, which help
the mixing of the IGM. These faint features are very 
difficult to detect, but seen
in a few systems, most notably
NGC 5044 (\citealt{buo03,dav09,dav10}). \\

iv) Cold and hybrid accretion models show short
quiescent periods (summing to a $\sim15\%$ of the evolutionary time).
Each quiescent period is followed by a relatively strong outburst,
with power about an order of magnitude larger than the typical
value during the continuous phase.
The consequence is the creation of relatively large
cavities ($10-20$ kpc), 
with high density contrast, in approximate pressure equilibrium
with the ambient. The emission-weighted temperature in the cavity
region, however, is not far above those
of the surrounding medium.
Feeble rims, relatively cold in the region near the center and
slightly hotter in the more distant part, with somewhat higher
iron abundance, are also present. These features are again
reminescent of real systems (\citealt{mcn07} for a review,
\citealt{gast09,dav09}).\\

v) The asymmetrical transport of metals, along jet-axis, up to
40 kpc, is clearly visible during the active phases. In quiescent
periods, the moderate turbulence
promotes metal diffusion. Recent observational
data confirms
this behaviour (\citealt{kir09,kir11,dav10}).\\

vi) Finally, the global point of all our computations:
the only possible way to heat a galaxy group, in order
to suppress the grip of the cooling flow,
appears to be through a quasi-continuous
`delicate touch', while in massive clusters
rare and explosive strokes are mildly favoured
over the weaker Bondi regulation. 
This quite different feedback evolution might explain the common observed
discrepancies between the two cosmological actors.
Galaxy groups are indeed not purely scaled-down versions of galaxy clusters.

\section*{Acknowledgments}
The software used in this work was in part developed by the DOE
supported ASCI/Alliances Centre for Thermonuclear Flashes at the
University of Chicago.
We acknowledge the CINECA Award N. HP10BPTM62, 2011 for the availability 
of high performance computing resources and support.
Some simulations were also performed at NASA/Ames `Pleiades': we
would like to thank Pasquale Temi, as the principal support scientist
at the NASA base.

\bsp

\label{lastpage}

\end{document}